\documentclass[12pt]{iopart}

\usepackage[english]{babel}
\usepackage[dvips]{graphicx}
\usepackage{amstext,amssymb}
\usepackage[comma,square,sort]{natbib}
\usepackage{array}

\def\21{SU(2) $\otimes$ U(1) }
\newcommand{\newblock}{}
\newcommand{\eqref}[1]{(\ref{#1})}

\newcommand{\eg} {{\it e.g.}}
\newcommand{\ie} {{\it i.e.}}

\newcommand{\Eps} {\varepsilon}
\newcommand{\Epp} {\varepsilon'}

\newcommand{\TM}  {TM}
\newcommand{\TMs} {TMs}
\newcommand{\La}{|\mathbf{\Lambda}|}

\newcommand{\CL}   {C.L.}
\newcommand{\dof}  {d.o.f.}
\newcommand{\eVq}  {\text{eV}^2}
\newcommand{\Sol}  {\textsc{sol}}
\newcommand{\SlKm} {\textsc{sol+kam}}
\newcommand{\Atm}  {\textsc{atm}}
\newcommand{\KtK}  {\textsc{k2k}}

\newcommand{\Dms}  {\Delta m^2_\Sol}
\newcommand{\Dma}  {\Delta m^2_\Atm}

\newcommand{\Lsnd} {\textsc{lsnd}}
\newcommand{\Dml}  {\Delta m^2_\Lsnd}
\newcommand{\Sbl}  {\textsc{sbl}}
\newcommand{\Nev}  {\textsc{nev}}

\begin{document}



\title{Status of global fits to neutrino oscillations}

\author{Michele Maltoni\dag, Thomas Schwetz\ddag, Mariam
  T{\'o}rtola\S\ and Jos{\'e} W.~F.~Valle\S}

\address{\dag\ C.N.~Yang Institute for Theoretical Physics, SUNY at
  Stony Brook, \\
  Stony Brook, NY 11794-3840, USA}

\address{\ddag\ Physik--Department, Technische Universit{\"a}t
  M{\"u}nchen, James--Franck--Strasse, \\
  D--85748 Garching, Germany}

\address{\S\ AHEP Group, Instituto de F\'{\i}sica Corpuscular --
  C.S.I.C./Universitat de Val{\`e}ncia, \\
  Edificio Institutos de Paterna, Apt 22085, E--46071 Valencia, Spain}

\ead{maltoni@insti.physics.sunysb.edu, schwetz@ph.tum.de,
  mariam@ific.uv.es, valle@ific.uv.es}

\begin{abstract}
    We review the present status of global analyses of neutrino
    oscillations, taking into account the most recent neutrino data
    including the latest KamLAND and K2K updates presented at
    Neutrino2004, as well as state-of-the-art solar and atmospheric
    neutrino flux calculations. We give the two-neutrino solar +
    KamLAND results, and the two-neutrino atmospheric + K2K
    oscillation regions, discussing in each case the robustness of the
    oscillation interpretation against departures from the Standard
    Solar Model and the possible existence of non-standard neutrino
    physics.
    Furthermore, we give the best fit values and allowed ranges of the
    three--flavour oscillation parameters from the current worlds'
    global neutrino data sample and discuss in detail the status of the
    small parameters $\alpha \equiv \Dms/\Dma$ as well as
    $\sin^2\theta_{13}$, which characterize the strength of CP violating
    effects in neutrino oscillations. We also update the degree of
    rejection of four--neutrino interpretations of the LSND evidence in
    view of the most recent developments.
\end{abstract}


\maketitle

\section{Introduction}
\label{sec:introduction}

The discovery of neutrino masses by the combination of a variety of
data from
solar~\cite{davis:1994jw,cleveland:1998nv,abdurashitov:1999zd,cattadori:2002rd,hampel:1998xg,altmann:2000ft,fukuda:2002pe,ahmad:2002jz,ahmad:2002ka,Ahmed:2003kj},
atmospheric~\cite{fukuda:1998mi,surdo:2002rk,Sanchez:2003rb},
reactor~\cite{eguchi:2002dm,araki:2004mb} and
accelerator~\cite{ahn:2002up,k2k-nu04} neutrino experiments was the
major recent achievement in astroparticle, high energy, and nuclear
physics, which culminates a heroic effort dating back to about four
decades. This has now firmly established the incompleteness of the
Standard Model of electroweak interactions, expected on theoretical
grounds since long
ago~\cite{gell-mann:1980vs,yanagida:1979,schechter:1980gr,mohapatra:1981yp,Valle:2003rh}.
The determination of neutrino oscillation parameters is now a
flourishing industry~\cite{industry} which has finally entered the
high precision age, with many experiments underway and a new
generation coming. Apart from a careful understanding of solar and
atmospheric neutrino fluxes, nuclear physics, neutrino cross sections
and experimental response functions, the interpretation of the data
relies heavily on the proper description of neutrino propagation
properties both in the Sun and the Earth, including the so--called
matter effects~\cite{mikheev:1985gs,wolfenstein:1978ue}.

The avalanche of data in a field where these have been traditionally
so scarce, given the feebleness of neutrino interactions, has prompted
a rush of phenomenological papers on the interpretation of neutrino
data. A number of reviews are already in the
market~\cite{pakvasa:2003zv,Barger:2003qi,gonzalez-garcia:2002dz,Fogli:2003eh}.
Here we revisit the latest global analysis of neutrino oscillation
parameters presented in Ref.~\cite{Maltoni:2003da} in view of the most
recent solar~\cite{Bahcall:2004fg} and atmospheric~\cite{Honda:2004yz}
neutrino flux calculations and of the new data presented at Neutrino
2004.

This paper is organized as follows.  In Sec.~\ref{sec:dominant32} we
briefly discuss the analysis of the atmospheric neutrino data
including the state-of-the-art three--dimensional calculation of the
atmospheric neutrino flux given in Ref.~\cite{Honda:2004yz}.  In order
to obtain the allowed ranges for the oscillation parameters $\Dma$ and
$\theta_\Atm$ we combine the atmospheric neutrino data from the
Super-K experiment with the accelerator data from the K2K experiment,
which provides the first independent confirmation of the oscillation
evidence from atmospheric neutrinos. We give also a short description
of robustness of the atmospheric data with respect to non-standard
neutrino interactions in Sec.~\ref{sec:robustn-oscill-inter}.

In Sec.~\ref{sec:dominant21} we briefly describe the solar neutrino
experiments and the KamLAND reactor neutrino experiment, and discuss
the two-neutrino interpretation of the data described by $\Dms$ and
$\theta_\Sol$. We comment on the implications of the recent update of
the Standard Solar Model (SSM) neutrino fluxes from
Ref.~\cite{Bahcall:2004fg}. Furthermore, in
Sec.~\ref{sec:robustn-oscill-inter-1} we discuss the robustness of the
oscillation interpretation of solar neutrino data when going beyond
the SSM framework or invoking non-standard neutrino properties. As
examples we consider the effects of solar radiative-zone density
fluctuations, convective-zone magnetic fields and the prospects of
probing electromagnetic neutrino properties with current and future
experiments.

After discussing the dominant oscillations in the two-flavour
approximation in Secs.~\ref{sec:dominant32} and \ref{sec:dominant21}
we devote Sec.~\ref{sec:three-flav-neutr} to the global three--neutrino
analysis of the data, combining all current neutrino oscillation data
except LSND. In addition to presenting the best fit values and allowed
ranges for the oscillation parameters, we update the status of the
three--flavour parameters $\alpha \equiv \Dms/\Dma$ and $\theta_{13}$
in view of the new data reported at Neutrino 2004.  In
Sec.~\ref{sec:four-neutr-oscill} we give an update of the status of
attempts to account for the LSND data in terms of four--neutrino
oscillations.
Finally we conclude and summarize our results in
Sec.~\ref{sec:summary-conclusions}. A detailed discussion of the
recent KamLAND results is given in \ref{app:kamland}.

\section{Leading oscillations with $\Dma$}
\label{sec:dominant32}

\subsection{Atmospheric neutrino oscillations}
\label{sec:atmosph-neutr-oscill}

In 1998 the Super-K experiment obtained evidence for neutrino
oscillations~\cite{fukuda:1998mi} from the observation of the zenith
angle dependence of their $\mu$-like atmospheric neutrino data. This
effect has been confirmed also by other atmospheric neutrino
experiments, see \eg\ Refs.~\cite{surdo:2002rk,Sanchez:2003rb}.
Recently, Super-K has reported a dip in the $L/E$ distribution of the
atmospheric $\nu_\mu$ survival probability~\cite{Ashie:2004mr}, which
provides a clear signature for neutrino oscillations.

In our atmospheric neutrino analysis we include the most recent
charged-current atmospheric neutrino data from
Super-K~\cite{hayato:2003}, including the $e$-like and $\mu$-like data
samples of sub- and multi-GeV contained events (each grouped into 10
bins in zenith angle), as well as the stopping (5 angular bins) and
through-going (10 angular bins) up-going muon data events.  As
previously, we do not use the information on $\nu_\tau$ appearance,
multi-ring $\mu$ and neutral-current events since an efficient
Monte-Carlo simulation of these data would require a more detailed
knowledge of the Super-K experiment, and in particular of the way the
neutral-current signal is extracted from the data. For details of our
analysis see Refs.~\cite{gonzalez-garcia:2000sq,maltoni:2002ni} and
references therein.
With respect to our previous atmospheric neutrino
analysis~\cite{Maltoni:2003da} we have now taken into account the new
three--dimensional atmospheric neutrino fluxes given in
Ref.~\cite{Honda:2004yz}. Furthermore, we have updated our statistical
analysis following closely Ref.~\cite{kameda}, taking special care of
systematical errors, like uncertainties in the neutrino fluxes and
detection cross sections. Details on the $\chi^2$ analysis can also be
found in Ref.~\cite{Fogli:2003th}. With these updates our results are
in excellent agreement with the ones of the Super-K
collaboration~\cite{hayato:2003}.

\begin{figure}[t] \centering
    \includegraphics[height=11cm]{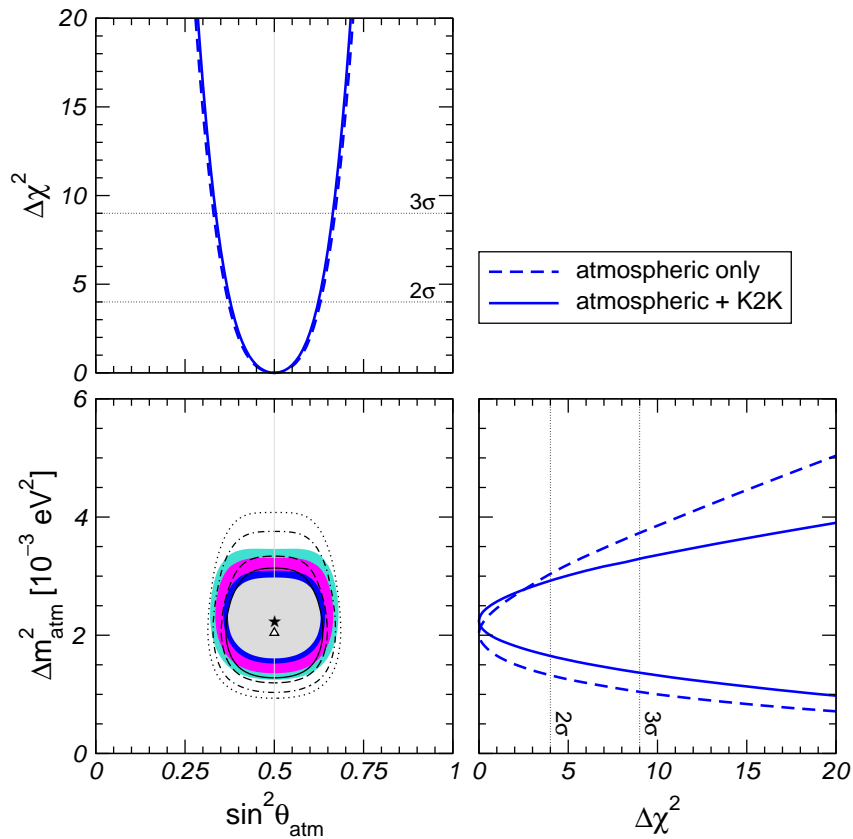}
    \caption{\label{fig:atm+k2k} %
      Allowed ($\sin^2\theta_\Atm$,~$\Dma$) regions at 90\%, 95\%,
      99\%, and 3$\sigma$ \CL\ for 2 \dof\ The regions delimited by
      the lines correspond to atmospheric data only, while for the
      colored regions also K2K data are added. The best fit point of
      atmospheric (atmospheric + K2K) data is marked by a triangle
      (star). Also shown is the $\Delta \chi^2$ as a function of
      $\sin^2\theta_\Atm$ and $\Dma$, minimized with respect to the
      undisplayed parameter.}
\end{figure}

In Fig.~\ref{fig:atm+k2k} we show the results of our analysis
of atmospheric data in the framework of two-flavour
$\nu_\mu\to\nu_\tau$ oscillations. The regions delimited by the hollow
contours correspond to the allowed regions for the
oscillation parameters $\sin^2\theta_\Atm$ and $\Dma$.
The current best fit point occurs at
\begin{equation}\label{eq:bfp-atm}
    \sin^2\theta_\Atm = 0.50 \,,\quad
    \Dma = 2.0 \times 10^{-3}~\eVq \quad\text{(ATM data)} \,.
\end{equation}
The main difference to the results of Ref.~\cite{Maltoni:2003da} are
the relatively lower values of $\Dma$ implied by the use of the new
three--dimensional atmospheric neutrino fluxes reported in
Ref.~\cite{Honda:2004yz} instead of the one--dimensional Bartol
fluxes~\cite{barr:1989ru} used previously.

\subsection{The K2K accelerator experiment}
\label{sec:k2k-accel-exper}

The KEK to Kamioka (K2K) long-baseline neutrino oscillation
experiment~\cite{ahn:2002up} probes the $\nu_\mu$ disappearance
oscillation channel in the same region of $\Delta m^2$ as explored by
atmospheric neutrinos. The neutrino beam is produced by a 12~GeV
proton beam from the KEK proton synchrotron, and consists of 98\% muon
neutrinos with a mean energy of 1.3~GeV. The beam is controlled by a
near detector 300~m away from the proton target. Information on
neutrino oscillations is obtained by the comparison of this near
detector data with the $\nu_\mu$ content of the beam observed by the
Super-Kamiokande detector at a distance of 250~km.

The data sample called K2K-I~\cite{ahn:2002up} has been collected in
the period from June 1999 to July 2001 ($4.8\times 10^{19}$ protons on
target). K2K-I has observed 56 events in Super-K, whereas
$80.1^{+6.2}_{-5.4}$ have been expected in the case of no
oscillations. This gives a clear evidence for $\nu_\mu$ disappearance:
the probability that the observed flux at Super-K is explained by a
statistical fluctuation without neutrino oscillations is less than
1\%~\cite{ahn:2002up}. Recently, at the Neutrino2004 conference new
data from the K2K-II period have been presented~\cite{k2k-nu04}. K2K-II
started in fall 2002, and the released data corresponds to $4.1\times
10^{19}$ protons on target, comparable to the K2K-I sample. From the
combined analysis of K2K-I and K2K-II 108 events have been observed in
Super-K, whereas $150.9^{+11.6}_{-10.0}$ have been expected for no
oscillations. Out of the 108 events 56 are so-called single-ring muon
events.  This data sample contains mainly muon events from the
quasi-elastic scattering $\nu_\mu + p \to \mu + n$, and the
reconstructed energy is closely related to the true neutrino energy.
Hence these data can be used for a spectral analysis. Using the
Kolmogorov-Smirnov test the K2K collaboration finds that the observed
spectrum is consistent with the spectrum expected for no oscillation
only at a probability of 0.11\%, whereas the spectrum predicted by the
best fit oscillation parameters has a probability of
52\%~\cite{k2k-nu04}.

\begin{figure}[t] \centering
    \includegraphics[height=9cm]{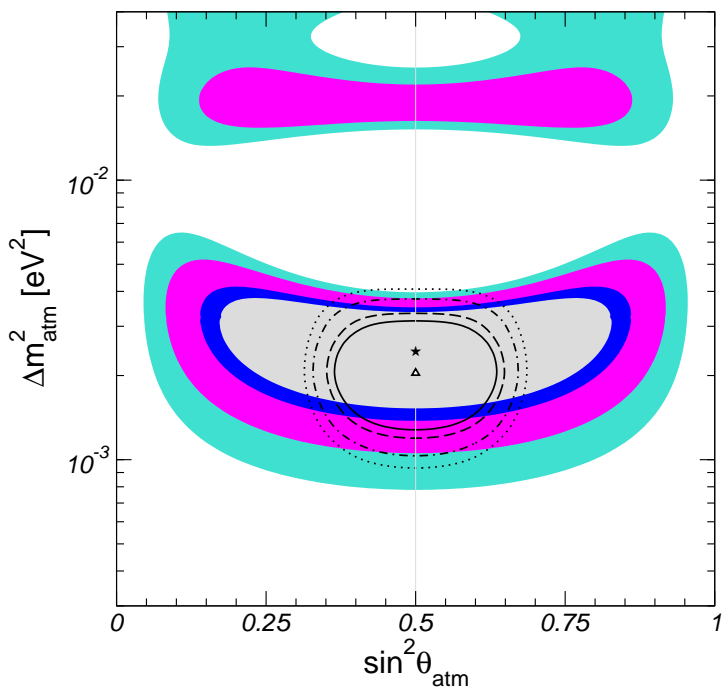}
    \caption{\label{fig:k2k} %
      Allowed K2K regions in the ($\sin^2\theta_\Atm$,~$\Dma$) plane
      at 90\%, 95\%, 99\%, and 3$\sigma$ \CL\ for 2 \dof\ The hollow
      lines delimit the region determined from the atmospheric data
      only. The star (triangle) corresponds to the K2K (atmospheric)
      best fit point.}
\end{figure}

In our re-analysis of K2K data we use the energy spectrum of the 56
single-ring muon events from K2K-I + K2K-II~\footnote{We cannot use
the full K2K data sample of 108 events, since not enough information
is available to analyze these data outside the K2K
collaboration.}. Similar as in Ref.~\cite{Fogli:2003th} we use a
phenomenological parameterization for the spectrum expected for no
oscillation. Adopting reasonable assumptions on the energy resolution
function and on systematical errors, we fit the data divided into 15
bins in reconstructed neutrino energy, as given in Ref.~\cite{k2k-nu04}.
The allowed regions for the oscillation parameters from K2K data are
shown in Fig.~\ref{fig:k2k} in comparison to the ones from atmospheric
neutrino data. This figure illustrates that the neutrino mass-squared
difference indicated by the $\nu_\mu$ disappearance observed in K2K is
in perfect agreement with atmospheric neutrino oscillations. Hence,
K2K data provide the first confirmation of oscillations with $\Dma$
from a man-made neutrino source. K2K gives a rather weak constraint on
the mixing angle due to low statistics in the current data sample.

\subsection{Atmospheric and K2K combined}
\label{sec:atmosph-k2k-comb}

In Fig.~\ref{fig:atm+k2k} the allowed regions in the
($\sin^2\theta_\Atm$,~$\Dma$) plane from the combined analysis of K2K
and Super-K atmospheric neutrino data are shown as shaded regions. As
expected from Fig.~\ref{fig:k2k} we find that, apart from providing an
independent confirmation, K2K data start already to constrain the
allowed region of $\Dma$, whereas the determination of the mixing
angle is completely dominated by atmospheric data.
From the projections of the $\chi^2$ onto the $\Dma$ and
$\sin^2\theta_\Atm$ axes shown in Fig.~\ref{fig:atm+k2k} we find the
best fit point
\begin{equation}\label{eq:bfp-atm+k2k}
    \sin^2\theta_\Atm = 0.5 \,,\quad
    \Dma = 2.2\times 10^{-3}~\eVq \quad\text{(ATM+K2K data)}
\end{equation}
with the corresponding allowed ranges at 3$\sigma$ ($5\sigma$) for 1
\dof:
\begin{eqnarray} \label{eq:atm_ranges}
    0.34~(0.27) \le \sin^2\theta_\Atm \le 0.66~(0.73) \,, \\
    1.4~(0.85)\times 10^{-3}~\eVq \le \Dma \le 3.3~(4.2)\times 10^{-3}~\eVq\,.
    \nonumber
\end{eqnarray}
Note that despite the downward shift of the atmospheric mass-splitting
implied by the new neutrino fluxes from Ref.~\cite{Honda:2004yz} our
new quoted value for $\Dma$ in Eq.~\eqref{eq:bfp-atm+k2k} is
statistically compatible both with our previous
result~\cite{Maltoni:2003da} and the value obtained by the new Super-K
$L/E$ analysis~\cite{Ashie:2004mr}. Let us remark that the K2K
constraint on $\Dma$ from below is important for future long-baseline
experiments, since the performance of such experiments is drastically
affected if $\Dma$ were in the lower part of the 3$\sigma$ range
allowed by atmospheric data (see \eg\ Ref.~\cite{Huber:2004ug}).

\subsection{Robustness of oscillation interpretation: atmospheric neutrinos}
\label{sec:robustn-oscill-inter}

We now turn to the issue of the robustness of the oscillation
interpretation of the atmospheric neutrino data. Non-standard physics
may in principle affect atmospheric neutrino fluxes, as well as
neutrino propagation and detection cross
sections~\cite{pakvasa:2003zv}.
Apart from the issue of their theoretical
viability~\cite{schechter:1982cv} neutrino decays have been considered
since long ago~\cite{pakvasa:1972gz}.  Non-standard interactions
arising from the \21 charged and neutral
currents~\cite{schechter:1980gr} or from new
particles~\cite{zee:1980ai,babu:1988ki,mohapatra:1986bd,hall:1986dx,bernabeu:1987gr,branco:1989bn,rius:1990gk},
as well as quantum mechanical decoherence~\cite{Fogli:2003th} might
also affect atmospheric neutrino results.  Although strongly rejected
by recent atmospheric data as the dominant
mechanisms~\cite{Ashie:2004mr,fornengo:2001pm,Fogli:2003th},
non-standard phenomena might still be present at a sub-leading level
in addition to oscillations and, to this extent, have some impact on
the determination of the oscillation parameters.

\begin{figure}[t] \centering
    \includegraphics[scale=.3]{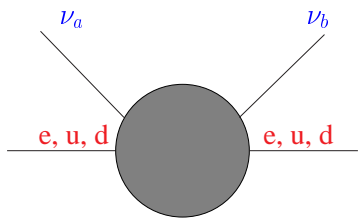}
    \caption{\label{fig:nuNSI} %
      Effective flavour-changing NSI operator.}
\end{figure}

In the following we illustrate the stability of the measurement of
$\Dma$ and $\sin^2\theta_\Atm$ by assuming the presence of
non-standard interactions of the neutrinos with earth matter.
New neutrino interactions beyond the Standard Model are a natural
feature in most neutrino mass models~\cite{valle:1991pk} and can be of
two types: flavour-changing (FC) and non-universal (NU).
These interactions (called NSI for short) may be schematically
represented as effective dimension-6 terms of the type $\varepsilon
G_F$, as illustrated in Fig.~\ref{fig:nuNSI}, where $\varepsilon$
specifies their sub-weak strength.
Such interactions may arise from a nontrivial structure of charged and
neutral current weak interactions characterized by a non-unitary lepton
mixing matrix~\cite{schechter:1980gr}. These gauge-induced NSI may
lead to flavour and CP violation, even with massless or degenerate
neutrinos~\cite{mohapatra:1986bd,bernabeu:1987gr,branco:1989bn,rius:1990gk}.
Alternatively, such non-standard neutrino interactions may also arise
in models where neutrino masses are ``calculable'' from radiative
corrections~\cite{zee:1980ai,babu:1988ki} and in some supersymmetric
models with broken R parity~\cite{Hirsch:2004he}.
Finally, in supersymmetric unified models, the strength of
non-standard neutrino interactions may be a calculable renormalization
effect~\cite{hall:1986dx}.

The impact of non-standard neutrino interactions on atmospheric
neutrinos was considered in Ref.~\cite{fornengo:2001pm} treating the
NSI strengths as free phenomenological parameters. This analysis takes
into account both the effect of $\nu_\mu \to \nu_\tau$ oscillations
(OSC) as well as the existence of non-standard neutrino--matter
interactions (NSI) in this channel. In addition to the standard term
in the Hamiltonian describing oscillations an term $H_\mathrm{NSI}$ is
introduced, accounting for an effective potential induced by the NSI
with earth matter:
\begin{equation}
    H_\mathrm{NSI} = \pm \sqrt{2} G_F N_f
    \left( \begin{array}{cc}
        0 & \varepsilon \\ \varepsilon & \varepsilon'
    \end{array}\right) \,.
\end{equation}
Here $+(-)$ holds for neutrinos (anti-neutrinos) and $\varepsilon$
and $\varepsilon'$ parameterize the NSI: $\sqrt{2} G_F N_f \varepsilon$
is the forward scattering amplitude for the FC process $\nu_\mu + f
\to \nu_\tau + f$ and $\sqrt{2} G_F N_f \varepsilon'$ represents the
difference between $\nu_\mu + f$ and $\nu_\tau + f$ elastic forward
scattering. The quantity $N_f$ is the number density of the fermion
$f$ along the neutrino path. For definiteness we take for $f$ the
down-type quark.

\begin{figure}[t] \centering
    \includegraphics[height=11cm]{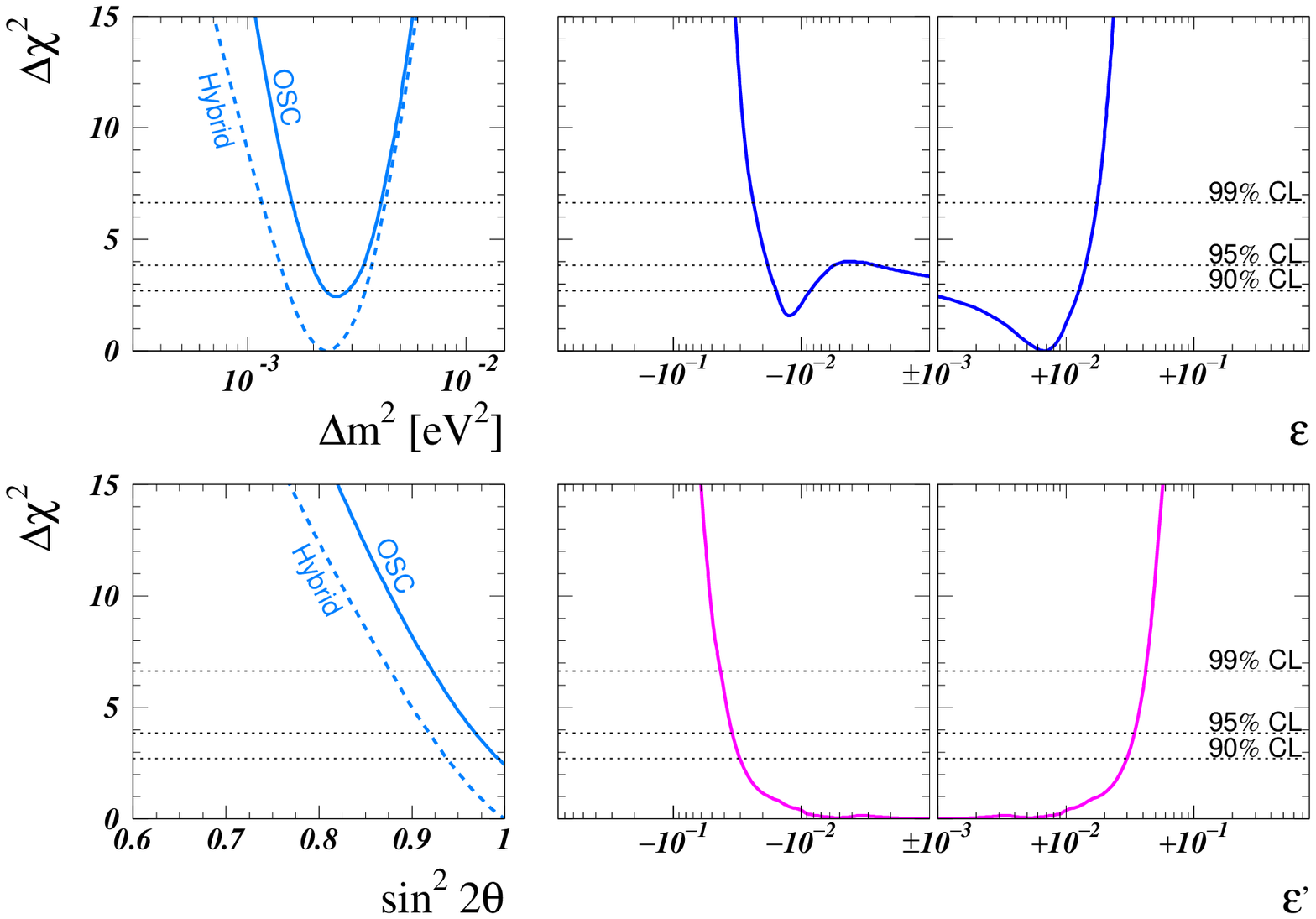}
    \caption{ \label{fig:chisq_eps} %
      Atmospheric neutrino oscillation parameters from
      ~\cite{fornengo:2001pm} fitted in the presence of non-standard
      interactions using the Bartol flux~\cite{barr:1989ru}. We show
      the behaviour of the $\chi^2$ as a function of the four
      parameters $\Dma$ (upper left), $\sin^2 2\theta_\Atm$ (lower
      left), the FC parameter $\Eps$ (upper right), and the NU
      parameter $\Epp$ (lower right).  In each panel the $\chi^2$ is
      minimized with respect to the three undisplayed parameters. In
      the left panels we show also the pure oscillation case (OSC),
      with $\Eps$ and $\Epp$ set to zero.}
\end{figure}

In Fig.~\ref{fig:chisq_eps} we show the results of a fit to
atmospheric neutrino data for the four parameters $\Dma$, $\sin^2
2\theta_\Atm$, $\Eps$, and $\Epp$~\footnote{Here we assume that the
  parameter $\Eps$ is real; for the more general case of complex
  $\Eps$, see Ref.~\cite{Gonzalez-Garcia:2004wg}.}. In the left panels
the pure oscillation case ($\Eps = \Epp =0$) is compared to the case
where also some NSI are allowed. We find that the $\chi^2$ improves
slightly (2.4 units) in the presence of NSI, but the determination of
the oscillation parameters is practically unaffected. This is an
important result, since it shows that the allowed ranges derived for
$\Dma$ and $\sin^2 2\theta_\Atm$ are rather stable with respect to
non-standard physics~\footnote{Note that the analysis of
  Ref.~\cite{fornengo:2001pm} is based on the neutrino fluxes of
  Ref.~\cite{barr:1989ru}, and in addition to the Super-K data
  described in Sec.~\ref{sec:atmosph-neutr-oscill} also the up-going
  muon data from the MACRO experiment~\cite{surdo:2002rk} is used.
  This explains the slight difference between the best fit values of
  $\Dma$ from Fig.~\ref{fig:chisq_eps} for the pure oscillation case
  and Eq.~\eqref{eq:bfp-atm}.}. In turn, the high preference of the
data for oscillations allows to set strong bounds on NSI. From the
right panels of Fig.~\ref{fig:chisq_eps} we deduce the bounds at
3$\sigma$
\begin{equation}
    -0.03 \le \Eps \le 0.02 \,,\quad |\Epp| \le 0.05 \,.
\end{equation}

Before closing we stress that, apart from its intrinsic theoretical
importance, the study of non-standard neutrino interactions has an
astrophysical interest, as these can affect the propagation of
neutrinos in a variety of astrophysical environments, such as
supernovae~\cite{nunokawa:1996tg,fogli:2002xj} and
pulsars~\cite{grasso:1998tt}. They could lead to ``deep-inside''
neutrino conversions, in addition to those expected from conventional
neutrino oscillations.


\section{Leading oscillations with $\Dms$}
\label{sec:dominant21}

\subsection{Solar neutrino oscillations}
\label{sec:solar-neutr-oscill}

In this section we consider oscillations of solar neutrinos in the
two-flavour framework. A detailed discussion of experimental and
theoretical aspects of solar neutrino physics can be found in other
contributions to this
volume~\cite{mcdonald,Bahcall:2004mz}. Therefore, we focus in the
following on the determination of the oscillation parameters from the
fit to solar neutrino data.
In our analysis we take into account the rates of the chlorine
experiment at the Homestake mine~\cite{cleveland:1998nv,davis:1994jw}
($2.56 \pm 0.16 \pm 0.16$~SNU), the most up-to-date
results~\cite{taup} of the gallium experiments
SAGE~\cite{abdurashitov:2002nt,abdurashitov:1999zd}
($66.9~^{+3.9}_{-3.8}~^{+3.6}_{-3.2}$~SNU) and
GALLEX/GNO~\cite{cattadori:2002rd,hampel:1998xg,altmann:2000ft} ($69.3
\pm 4.1 \pm 3.6$~SNU), as well as the 1496--day Super-K data
sample~\cite{fukuda:2002pe} in the form of 44 bins (8 energy bins, 6
of which are further divided into 7 zenith angle bins). From the SNO
experiment we include the most recent data from the salt
phase~\cite{Ahmed:2003kj} in the form of the neutral current (NC),
charged current (CC) and elastic scattering (ES) fluxes, as well as
the 2002 spectral day/night data~\cite{ahmad:2002jz,ahmad:2002ka} (17
energy bins for each day and night period).

The analysis methods used here are similar to the ones described in
Refs.~\cite{maltoni:2002ni,Maltoni:2003da} and references therein,
including the use of the so-called pull approach for the $\chi^2$
calculation, as described in Ref.~\cite{Fogli:2002pt}. In this method
all systematic uncertainties are included by introducing new
parameters in the fit and adding a penalty function to the $\chi^2$.
For example, for each of the eight solar neutrino fluxes a parameter
is introduced, and the predictions from the Standard Solar Model
including the correlated errors are taken into account by means of a
penalty function. The method described in Ref.~\cite{Fogli:2002pt} is
extended in two respects. First, it is generalized to the case of
correlated statistical errors~\cite{Balantekin:2003jm} as necessary to
treat the SNO--salt data. Second, we do not consider the $\chi^2$ only
up to first order in the pulls, but instead each pull parameter is
treated exactly to all orders. This is particularly interesting in the
case of the solar $^8$B flux. In our approach it is possible to
include the SSM prediction for this flux as well as the SNO NC
measurement on the same footing, without pre-selecting a particular
value, as implied by expanding around the predicted value. In this way
the fit itself can choose the best compromise between the SNO NC data
and the SSM prediction.

\begin{figure}[t] \centering
    \includegraphics[height=11cm]{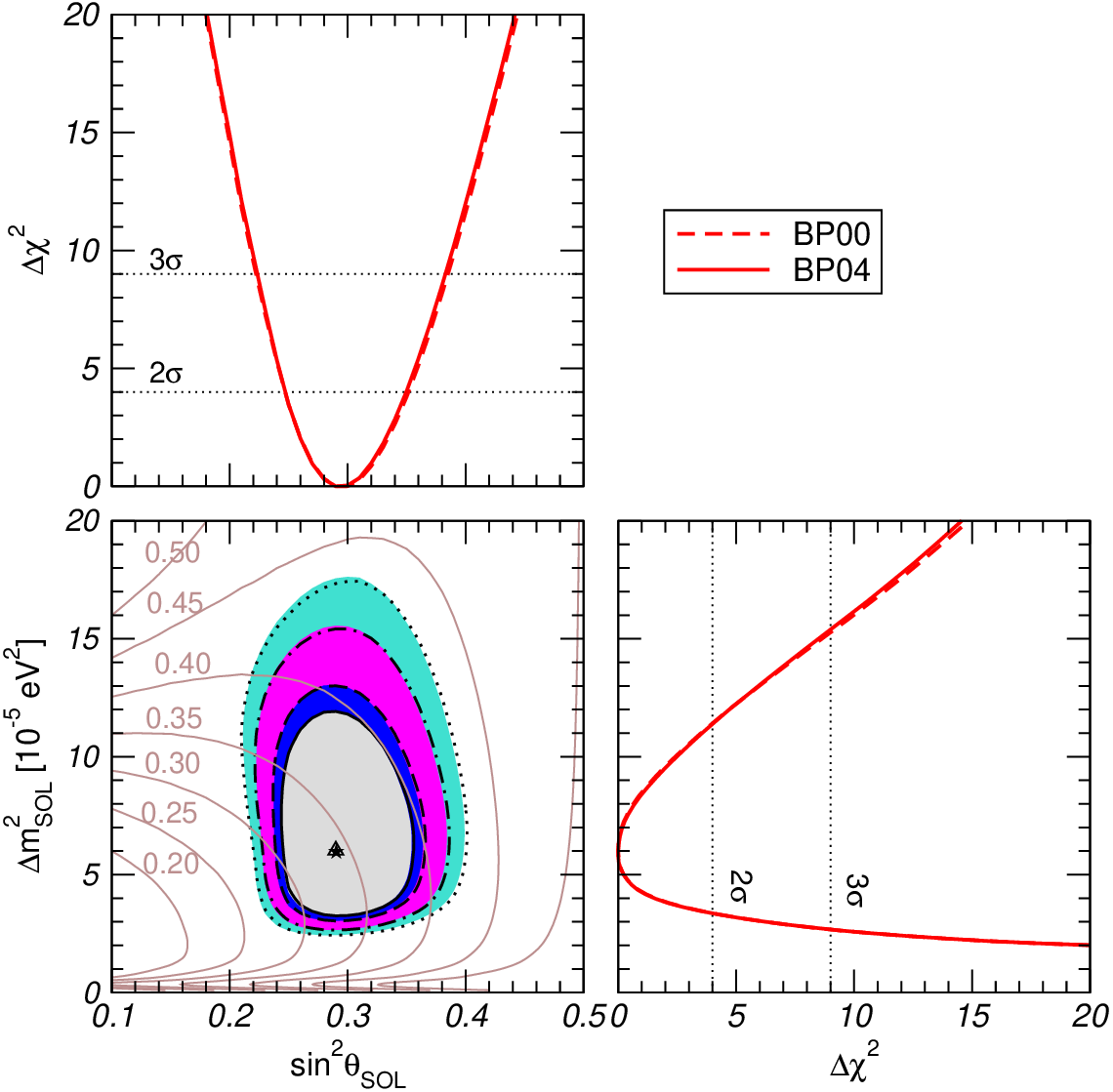}
    \caption{\label{fig:sol-region} %
      Allowed regions from all solar neutrino data at 90\%, 95\%,
      99\%, and 3$\sigma$ \CL\ for 2 \dof\ in the plane of
      $\sin^2\theta_\Sol$ and $\Dms$.  The regions delimited by the
      curves correspond to the BP00 SSM~\cite{Bahcall:2000nu}, whereas
      for the colored regions the BP04 SSM~\cite{Bahcall:2004fg} has
      been used.  Also shown is $\Delta \chi^2$ as a function of
      $\sin^2\theta_\Sol$ and $\Dms$, minimized with respect to the
      undisplayed parameter. The labeled contours denote constant
      CC/NC ratio in the SNO experiment.}
\end{figure}

In Fig.~\ref{fig:sol-region} we compare the allowed regions for the
oscillation parameters using the solar neutrino fluxes given in the
BP00 SSM~\cite{Bahcall:2000nu} and the recent update (BP04) presented
in Ref.~\cite{Bahcall:2004fg}. One finds that the change in the flux
predictions has a negligible impact on the allowed regions. This
illustrates that thanks to the good experimental accuracy the
determination of the oscillation parameters is rather robust with
respect to changes in the SSM. The current best fit values for solar
neutrino oscillation parameters are
\begin{equation}\label{eq:bfp-solar}
    \sin^2\theta_\Sol = 0.29 \,,\qquad
    \Dms = 6.0 \times 10^{-5}~\eVq \qquad\text{(solar data, BP04)}.
\end{equation}
Also the rejection against maximal solar mixing is 5.6$\sigma$, the
same as found previously~\cite{Maltoni:2003da} using the BP00 solar
model.  This is the significance at which bi--maximal models of
neutrino mass, such as the CP conserving version of the neutrino
unification model given in Ref.~\cite{chankowski:2000fp}, are ruled
out.  Despite the fact that the BP04 model does not imply any
significant change in the solar neutrino parameters, we will use it in
our subsequent discussion of solar neutrino results.


\subsection{The KamLAND reactor neutrino experiment}
\label{sec:kaml-react-neutr}

The KamLAND experiment~\cite{eguchi:2002dm,araki:2004mb} is a reactor
neutrino experiment with its detector located at the Kamiokande
site. Most of the $\bar{\nu}_e$ flux incident at KamLAND comes from
nuclear plants at distances of $80-350$ km from the detector, making
the average baseline of about 180 kilometers, long enough to provide a
sensitive probe of the LMA solution of the solar neutrino problem.
The KamLAND collaboration has for the first time measured the
disappearance of neutrinos traveling to a detector from a power
reactor.  They observe a strong evidence for the disappearance of
neutrinos during their flight over such distances, giving the first
terrestrial confirmation of the solar neutrino anomaly and also
establishing the oscillation hypothesis with man-produced neutrinos.

In KamLAND the reactor anti-neutrinos are observed by the process
$\bar\nu_e + p \to e^+ + n$, where the delayed coincidence of the
prompt energy from the positron and a characteristic gamma from the
neutron capture allows an efficient reduction of backgrounds. The
neutrino energy is related to the prompt energy by $E_\nu = E_\mathrm{pr} +
\Delta - m_e$, where $\Delta$ is the neutron-proton mass difference
and $m_e$ is the positron mass.
In the lower part of the energy spectrum there is a relevant
contribution from geo-neutrino events to the signal (see, \eg,
Refs.~\cite{Fiorentini:2003ww,Nunokawa:2003dd}). To avoid large
uncertainties associated with the geo-neutrino flux an energy cut at
2.6~MeV prompt energy is applied for the oscillation analysis.

First results from KamLAND were published in
Ref.~\cite{eguchi:2002dm}.  In the period from March to October 2002
data corresponding to a 162 ton-year exposure have been collected, and
after all cuts 54 anti-neutrino events remained in the final sample.
This number has to be compared with $86.8 \pm 5.6$ reactor neutrino
events predicted for no oscillations and $0.95\pm 0.99$ background
events, which gives a probability that the KamLAND result is
consistent with the no--disappearance hypothesis of less than
0.05\%~\cite{eguchi:2002dm}.
Recently, new results have been presented by
KamLAND~\cite{araki:2004mb}. With a somewhat larger fiducial volume of
the detector an exposure corresponding to 766.3~ton-year has been
obtained between March 2002 and January 2004 (including a reanalysis
of the 2002 data from Ref.~\cite{eguchi:2002dm}). A total of 258
events has been observed, in comparison to the expectation of
$356.2\pm 23.7$ reactor neutrino events in the case of no
disappearance and $7.5\pm 1.3$ background events. This leads to a
confidence level of 99.995\% for $\bar\nu_e$ disappearance, and the
averaged survival probability is $0.686 \pm 0.044\mathrm{(stat)} \pm
0.045\mathrm{(syst)}$. Moreover evidence for spectral distortion
consistent with oscillations is obtained~\cite{araki:2004mb}.

It was shown in Ref.~\cite{Schwetz:2003se} in relation with the first
KamLAND data that most information on the neutrino oscillation
parameters $\sin^2\theta_\Sol$ and $\Dms$ can be extracted from the
data by using an event-based likelihood analysis. This requires the
knowledge of the prompt energy of each observed event. Unlike to the
first data sample it is not possible to recover this information for
the latest KamLAND data sample from publicly available material.
Therefore only $\chi^2$-analyses based on binned data can be performed
outside the KamLAND collaboration.  Details of our KamLAND simulation
based on 2002 data can be found in
Refs.~\cite{maltoni:2002aw,Schwetz:2003se}, a discussion of our
updated analysis method for the current data sample is given in
\ref{app:kamland}. Here we briefly summarize the main features:
Instead of the traditional bins of equal size in $E_\mathrm{pr}$ we
use the data binned equally in $1/E_\mathrm{pr}$, which can be
obtained from Ref.~\cite{araki:2004mb}, and which allows to extract
more relevant information on the oscillation parameters.
We adopt the improved anti-neutrino flux parameterization from
Ref.~\cite{Huber:2004xh}, and include various systematic errors
associated to the neutrino fluxes, reactor fuel composition and
individual reactor powers.
Furthermore, we include matter effects, a careful treatment of
backgrounds and information on the average contribution to the total
reactor neutrino signal as a function of the distance to the detector.

\begin{figure}[t] \centering
    \includegraphics[height=9cm]{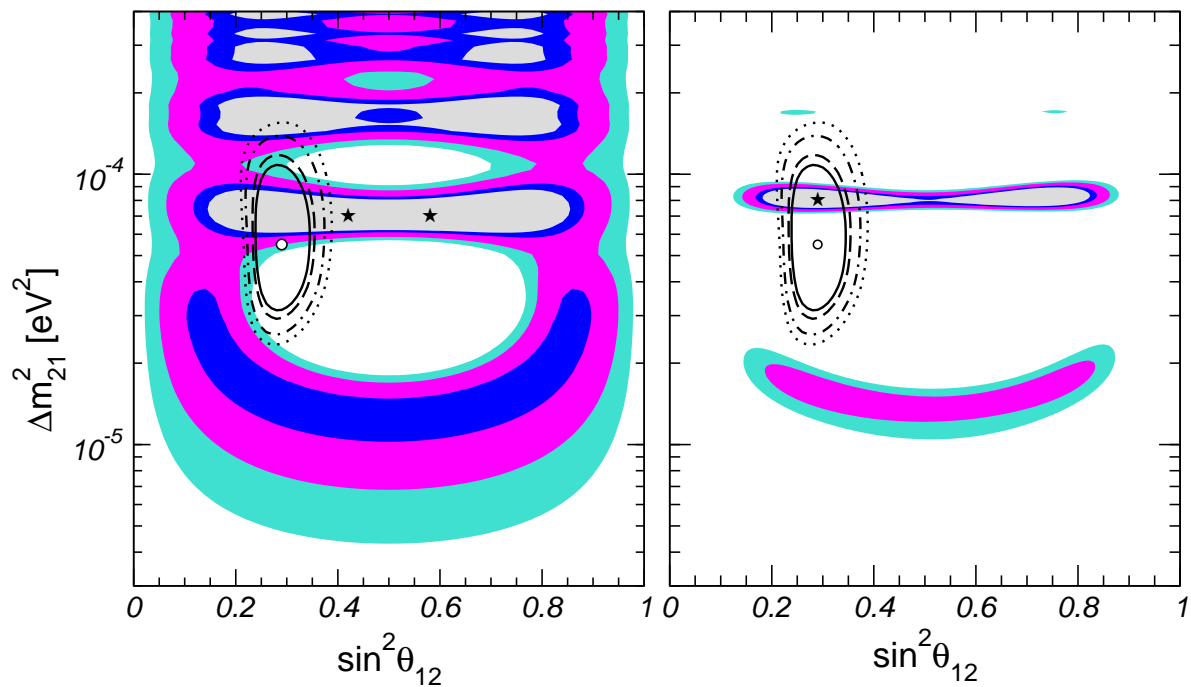}
    \caption{\label{fig:kaml-region} %
       Allowed regions from KamLAND data
       at 90\%, 95\%, 99\%, and 3$\sigma$ \CL\ for 2 \dof\ for the
       2002 data sample~\cite{eguchi:2002dm} (left panel) and the
       latest  766.3~ton-year data sample~\cite{araki:2004mb} (right panel). The
       regions delimited by the lines correspond to solar data. The
       KamLAND best fit points are marked with a star, the solar best
       fit point with a dot.}
\end{figure}

The KamLAND allowed regions for $\sin^2\theta_\Sol$ and $\Dms$ are
shown in Fig.~\ref{fig:kaml-region} in comparison to the regions from
solar data. One observes beautiful agreement between KamLAND data and
the region implied by the LMA solution to the solar neutrino problem,
which in this way has been singled out as the only viable one. Before
this experiment we had a very complex pattern of alternative
oscillation solutions like LOW, SMA, or VAC, see \eg,
Refs.~\cite{gonzalez-garcia:2000sq,maltoni:2002ni}. All of these are
completely ruled out by the KamLAND data. From this point of view the
KamLAND experiment has played a fundamental role in the resolution of
the solar neutrino problem.

Comparing the left and right panel of Fig.~\ref{fig:kaml-region} one
can appreciate the improvement implied by the recent KamLAND data. 
The allowed region is drastically reduced, and in particular the
mass-squared difference is very well determined by recent data. We
find the best fit point 
\begin{equation}
\begin{array}{l}    
\Dms = 8.1^{+0.4}_{-0.3} \times 10^{-5}~\eVq \,(1\sigma) \\
\sin^2 \theta_\Sol = 0.29
\end{array}
\qquad\text{(KamLAND data)}.
\end{equation}
The solution at low values of $\Dms$ is present at the 99\% \CL, with
the local minimum at $\Dms = 1.6 \times 10^{-5}~\eVq$ and $\sin^2
\theta_\Sol = 0.31$ and $\Delta \chi^2 = 7.5$ relative to the best fit
point. The so-called high-LMA solution is present only marginally at
the $3\sigma$ level at $\Dms = 1.7 \times 10^{-4}~\eVq$, $\sin^2
\theta_\Sol = 0.25$ and $\Delta \chi^2 = 11.3$. Although not
statistically significant, we note that the small matter effect
favours slightly values of $\sin^2\theta_\Sol < 0.5$ over the ``mirror
solution'' with $\sin^2\theta_\Sol > 0.5$.
The high values of $\Dms \gtrsim 3 \times 10^{-4}~\eVq$ in the
averaging regime, which have been allowed by first KamLAND
data~\cite{eguchi:2002dm} at 90\%~\CL\ are now ruled out at more than
3$\sigma$. This indicates that the spectral distortion associated to
neutrino oscillations has been observed~\cite{araki:2004mb} (see also
\ref{app:kamland} for further discussion).


\subsection{Solar and KamLAND combined}
\label{sec:solar-kaml-comb}

\begin{figure}[t] \centering
    \includegraphics[height=11cm]{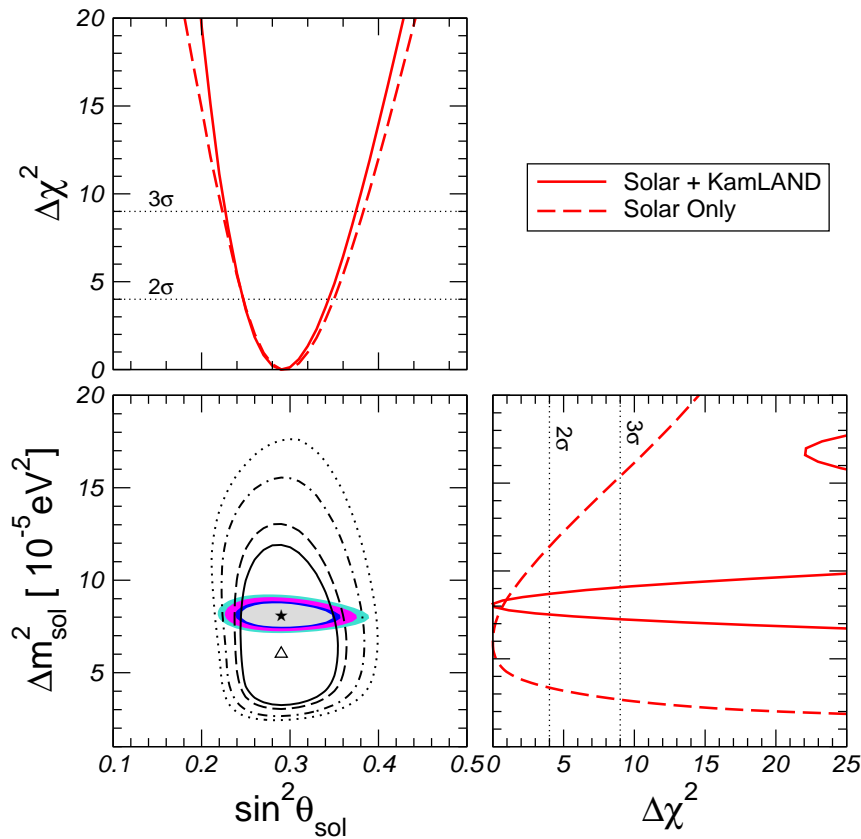}
    \caption{\label{fig:solkam-region} %
      Allowed regions from combined solar and KamLAND data at 90\%,
      95\%, 99\%, and 3$\sigma$ \CL\ for 2 \dof. The regions shown
      with lines correspond to solar data only, from
      Fig.~\ref{fig:sol-region}.}
\end{figure}

Under the fundamental assumption of CPT invariance we can directly
compare the information obtained from solar neutrino experiments and
the KamLAND reactor experiment.
In Fig.~\ref{fig:solkam-region} we show the allowed regions from the
combined solar and KamLAND data.  The current best fit point of the
global analysis occurs at
\begin{equation}\label{eq:bfp-sol+kaml}
    \sin^2\theta_\Sol = 0.29 \,,\quad
    \Dms = 8.1\times 10^{-5}~\eVq \quad\text{(Solar+KamLAND data)}.
\end{equation}
From the projections of the $\chi^2$ onto the $\Dms$ and
$\sin^2\theta_\Sol$ axes also shown in Fig.~\ref{fig:solkam-region} we
find the following allowed ranges at 3$\sigma$ ($5\sigma$) for 1 \dof:
\begin{eqnarray} \label{eq:solkam_ranges}
    0.23~(0.19) \le \sin^2\theta_\Sol \le 0.37~(0.45)\,, \\
    7.3~(6.7)\times 10^{-5}~\eVq \le \Dms  \le 9.1~(9.9)\times
    10^{-5}~\eVq\,.
    \nonumber
\end{eqnarray}
As expected from Fig.~\ref{fig:kaml-region} the determination of the
mixing angle is completely dominated by the solar neutrino data,
whereas KamLAND significantly reduces the allowed range for $\Dms$.
Also the so-called high-LMA region, which previously was present
around $\Dms\simeq 1.4 \times 10^{-4}~\eVq$ at $3\sigma$ (see \eg,
Ref.~\cite{Maltoni:2003da}) is now ruled out with a $\Delta\chi^2 =
22.1$ with respect to the global minimum, which corresponds to an
exclusion at about $4.3\sigma$ for 2~\dof.
Notice also that the day/night data are treated as
previously~\cite{maltoni:2002ni,Maltoni:2003da}. An improved analysis
of the day/night asymmetry data along the lines followed by the
Super-K collaboration in Ref.~\cite{Smy:2003jf} would lead
to an even more pronounced rejection of the high $\Dms$ region.  Currently
not enough information is available to reproduce this result outside
the Super-K collaboration.
Finally we note that the use of the BP04 fluxes has not changed
significantly the allowed regions.

\subsection{Robustness of oscillation interpretation: solar neutrinos}
\label{sec:robustn-oscill-inter-1}

The oscillation interpretation of solar neutrino data depends both on
astrophysical input (the model of the Sun) as well as on the physics
characterizing both the propagation as well as neutrino interaction
properties. Either may differ from the Standard Solar Model and
Standard Electroweak Model expectations.
How robust is the oscillation interpretation of solar neutrino data in
view of this?

\subsubsection{Beyond the Standard Solar Model}
\label{sec:beyond-stand-solar}

\begin{figure}[t] \centering
    \includegraphics[height=5cm]{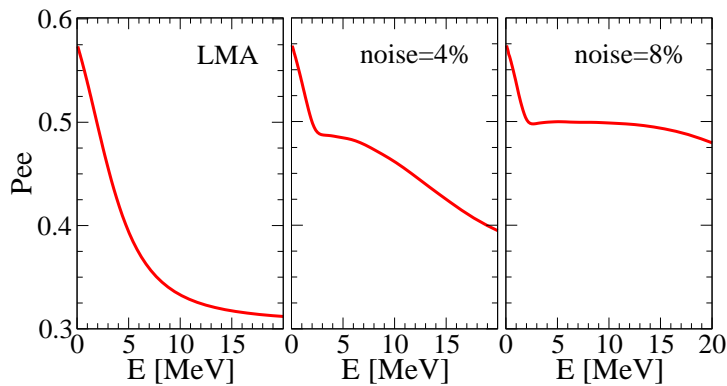}
    \caption{\label{fig:noisyLMA} %
      Effect of random matter density fluctuations with a correlation
      length of $L_0 = 100$~km on the electron-neutrino survival
      probability for LMA oscillations.}
\end{figure}

In the following we will briefly discuss consequences of departures
from the Standard Solar Model. The effect of varying solar neutrino
fluxes has been widely discussed, thus, as a case study we consider
the possibility of solar density fluctuations.
Even though this possibility was suggested in a number of
papers~\cite{balantekin:1996pp,nunokawa:1996qu,bamert:1998jj} it has
been traditionally neglected for several reasons. First, helioseismic
measurements constrain deviations of solar properties from Standard
Solar Model predictions at better than the percent level.  Second,
preliminary studies of the implications for neutrino oscillations of
radiative-zone helioseismic waves~\cite{bamert:1998jj} indicated that
they were unlikely to have observable effects.  Third, no other known
sources of fluctuations seemed to have the properties required to
influence neutrino oscillations.

Recently all of these points have been re-examined, with the result
that the presence of solar fluctuations appears more likely than
previously thought. First, direct helioseismic bounds turn out to be
insensitive to fluctuations whose size is as small as those to which
neutrinos are
sensitive~\cite{Castellani:1997pk,Christensen-Dalsgaard:2002ur}
typically several hundreds of km. Second, recent studies have shown
how such solar density fluctuations can arise near the solar
equatorial plane in the presence of magnetic fields deep within the
solar radiative zone due to a resonance between Alfv\'en waves and
helioseismic $g$-modes~\cite{Burgess:2003fj}.

\begin{figure}[t] \centering
    \includegraphics[height=7cm]{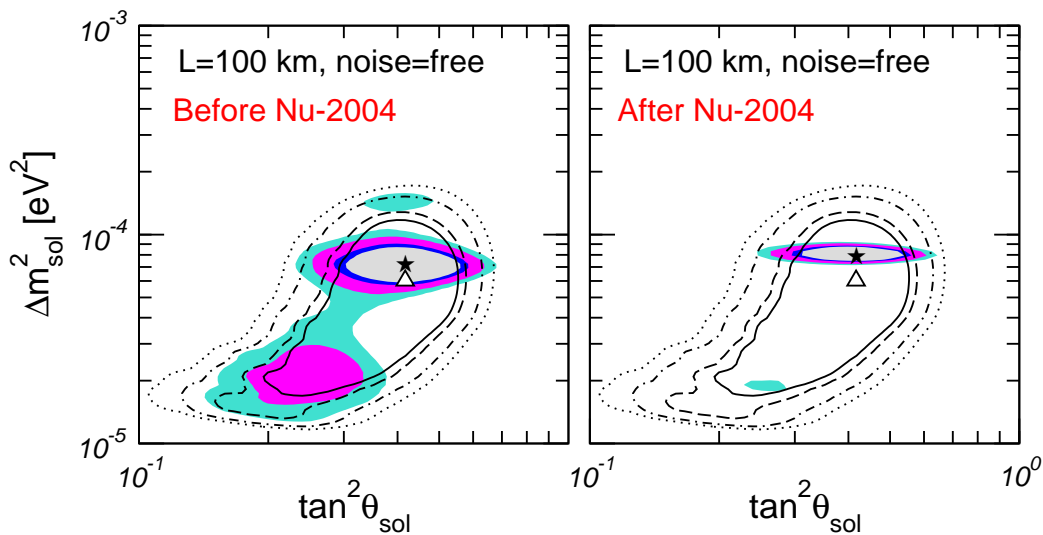}
    \caption{\label{fig:LMAwwonoise} Solar neutrino oscillation
      parameters with an arbitrary noise amplitude and a correlation
      length $L_0 = 100$~km from combining solar data with 2002
      KamLAND data (left panel) and 2004 KamLAND data (right panel). 
      The contour lines refer to solar data only.}
\end{figure}

It has been shown in Ref.~\cite{burgess:2002we} that such density
fluctuations can affect neutrino propagation in an important way.  The
effect of random matter density fluctuations on the electron-neutrino
survival probability for LMA oscillations has been shown to be
sizable if the correlation length $L_0$ (we take $L_0 = 100$ km) is
comparable to the neutrino oscillation length in the Sun. This is
illustrated in Fig.~\ref{fig:noisyLMA}.  The fluctuation's amplitude
$\xi$ at the position of neutrino resonance is zero in the left panel,
and is $\xi=4 \%$ and $\xi=8 \%$ in the middle and right panels,
respectively.
The corresponding solar neutrino oscillation parameters obtained in
our global fit are shown in Fig.~\ref{fig:LMAwwonoise} before and
after the new KamLAND data presented at Neutrino 2004.
One sees that these new data have a rather strong impact on the
stability of the oscillation parameters.  With 2002 KamLAND data an
additional allowed region was present for $\Dms \sim 2\times
10^{-5}~\eVq$ at the 99\% \CL~\cite{Burgess:2003su}, whereas the new
KamLAND data pin down the oscillation parameters such that the
allowed regions in Fig.~\ref{fig:LMAwwonoise} are practically stable,
and the new solar noise-induced solution appears only marginally at
the $3\sigma$ confidence level.

Conversely, as shown in Ref.~\cite{Burgess:2003su}, the quality of
current solar neutrino measurements after the SNO-salt and KamLAND
results is sufficiently good as to place important constraints on
fluctuations in the solar medium deep within the solar radiative zone.
In other words, neutrinos may be used as an astrophysical probe of the
solar interior, beyond the framework of the Standard Solar Model.
\begin{figure}[t] \centering
    \includegraphics[height=5.5cm]{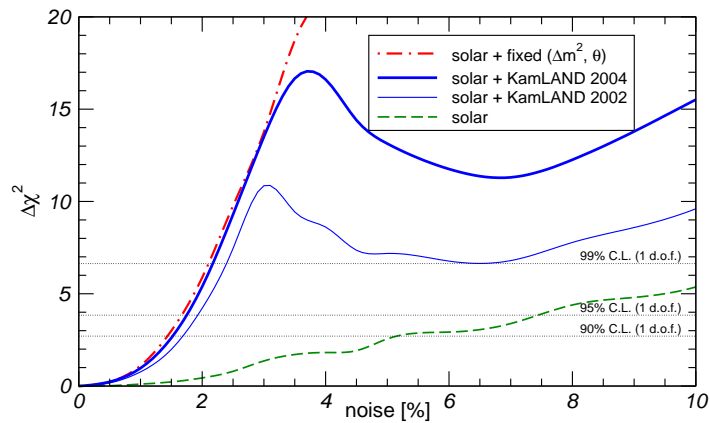}
    \caption{\label{fig:noiseBound} Bounds on random matter density
      fluctuations for a correlation length of $L_0 = 100$~km from
      solar neutrino data, solar + KamLAND data, and solar data for
      oscillation parameters fixed at the best fit point.}
\end{figure}
As illustrated in Fig.~\ref{fig:noiseBound} density fluctuations are
strongly constrained if the correlation length lies in the range of
several hundred km. Comparing the curves for free and fixed
oscillation parameters one notes that the bounds on fluctuations have
already become rather stable, a situation that may still improve when
the accuracy on $\Dms$ and $\theta_\Sol$ gets improved by future
neutrino experiments.
Because oscillations are sensitive to correlation lengths which are so
short, such solar neutrino results will complement the constraints
that come from helioseismology.

 \subsubsection{Beyond solar neutrino oscillations: Spin Flavour
   Precession}
 
 In extensions of the Standard Model (SM) neutrino masses are in
 general accompanied also by non-standard interactions and/or
 electromagnetic properties.  In the minimal extension of the Standard
 Model with Dirac neutrino masses one expects very tiny neutrino
 magnetic moments (MMs)~\cite{fujikawa:1980yx}, well below current
 experimental sensitivities.  However, the theoretically preferred
 case of Majorana neutrinos leads to potentially larger transition
 magnetic moments closer to the present sensitivities.
 These can affect neutrino propagation properties in the Sun beyond
 the oscillation mechanism, due to the possible presence of solar
 magnetic fields. Alternatively, they can affect the determination of
 neutrino oscillation parameters due to non-standard neutrino cross
 sections inside the detectors.
 
 The most general form of the electromagnetic current of massive
 (Majorana) neutrinos has been given in Ref.~\cite{Schechter:1981hw}.
 The magnetic piece is characterized by a $3 \times 3$ complex
 antisymmetric matrix, the so-called Majorana transition moment (\TM)
 matrix, that contains MMs as well as electric dipole moments of the
 neutrinos.
 Their existence would affect neutrino propagation inside the solar
 convective zone due to a spin-flavour precession (SFP)
 effect~\cite{Schechter:1981hw,akhmedov:1988uk,lim:1988tk}.  This in
 general depends on the assumed magnetic field profile. 
 In order to quantify the extent with which the oscillation regions
 can be altered by the sub-leading spin-flavour precession effect, a
 $\chi^2$ analysis was performed in Ref.~\cite{Miranda:2003yh} taking
 into account the global solar + KamLAND disappearance data. We
 assumed that neutrino conversions are driven mainly by LMA
 oscillations, and used the same self-consistent~\cite{Kutvitskii}
 convective--zone solar magnetic field profile employed
 in Ref.~\cite{Miranda:2000bi}.
 The results we obtain indicate that, even though small, current
 bounds on neutrino magnetic moments and solar magnetic fields still
 leave room for slight modifications in the determinations of solar
 neutrino oscillation parameters, in the presence of large magnetic
 moments~\footnote{Note that our analysis of solar neutrino data
   applies also to the special case of Dirac neutrinos.}.
 
 However, in the general Majorana case, where theory may give rise to
 higher moments, there is a characteristic feature of the spin flavour
 precession which will lead to more stringent constraints, and hence
 increase the robustness of the oscillation parameter determination.
 The argument is based on the presence of anti-neutrinos in the solar
 flux~\cite{Schechter:1981hw,akhmedov:1988uk,lim:1988tk}.  Recently
 the KamLAND collaboration~\cite{Eguchi:2003gg} has reported a result
 which greatly improves the bound on an anti-neutrino component in the
 solar flux from 0.1\% of the solar boron $ \nu_e $ flux to $
 2.8\times 10^{-2}$\% at the 90\%~\CL, about 30 times better than the
 recent Super-K limit~\cite{Gando:2002ub}. This implies that, in
 practical terms, extremely good stability of the solar neutrino
 oscillation parameters against the possible existence of sub-leading
 SFP conversions is obtained.  As a result, we conclude that solar
 neutrinos oscillation parameters can be inferred without any
 reference to neutrino magnetic properties nor solar magnetic fields.
 We refer the reader to Ref.~\cite{Miranda:2003yh} for quantitative
 details.
 
 All in all, our analysis of solar neutrino data implies indicates
 pretty good stability of the oscillation parameter determination for
 the case of Majorana neutrinos, due to the solar anti--neutrino limit
 from KamLAND.  In contrast for the special case of Dirac neutrinos
 this limit does not apply and the determination of oscillation
 parameters is potentially more fragile.  However, we note that the
 gauge theoretic expectations for Dirac magnetic moments are typically
 lower than those for Majorana neutrino transition moments.

\subsubsection{Beyond SM neutrino cross sections: Constraining
  neutrino magnetic moments}

Neutrino transition magnetic moments are basic properties of
neutrinos~\cite{Schechter:1981hw}. Although they do not substantially
affect neutrino propagation, even in the presence of solar magnetic
fields, non-trivial electromagnetic neutrino properties could still
show up in the detection process and to this extent affect the
determination of oscillation parameters.
Experiments based on the neutrino detection via neutrino--electron
elastic scattering are a sensitive probe of the electromagnetic
properties. In Ref.~\cite{grimus:2002vb} it was shown that current
data from solar neutrinos (in particular from Super-K) in combination with
reactor neutrino--electron scattering data provides strong bounds on
all the elements of the \TM\ matrix (for similar analyses see
Ref.~\cite{beacom:1999wx,Joshipura:2002bp,Grifols:2004yn,Liu:2004ny}).

In several experiments such as Super-K, Borexino and some reactor
experiments~\cite{derbin:1994ua,li:2002pn,Daraktchieva:2003dr},
neutrinos are detected via the elastic neutrino--electron scattering,
whose electromagnetic cross section
is~\cite{Bardin:1970wr,Kyuldjiev:1984kz}
\begin{equation}\label{eq:cross}
    \frac{d\sigma_\mathrm{em}}{dT} = \frac{\alpha^2 \pi}{m_e^2 \mu_B^2}
    \left(\frac{1}{T} - \frac{1}{E_\nu}\right) \mu^2_\mathrm{eff} \,,
\end{equation}
where $\mu_\mathrm{eff}$ is an effective MM~\cite{grimus:2000tq}, $T$
denotes the kinetic energy of the recoil electron and $E_\nu$ is the
incoming neutrino energy.
The electromagnetic cross section adds to the weak cross section and
allows to extract information on the \TM\ matrix, which we denote by
$\lambda$ in the following. Taking into account the antisymmetry of
$\lambda$ for Majorana neutrinos, it is useful to define vectors
$\mathbf\Lambda$ by $\lambda_{jk} = \varepsilon_{jkl} \Lambda_l$,
where $\lambda_{jk}$ are the elements of the \TM\ matrix in the mass
basis.
The effective MM square $\mu_\mathrm{eff}^2$ takes on different forms
in the cases of solar and reactor neutrino experiments.  The detailed
derivation of the following expressions can be found in
Ref.~\cite{grimus:2002vb}. For the case of solar neutrino experiments
one obtains the effective MM square
\begin{equation}\label{finalLMA}
    \mu^2_\mathrm{LMA} = \La^2 - |\Lambda_2|^2 +
    P^{2\nu}_{e1} \left( |\Lambda_2|^2 - |\Lambda_1|^2 \right) \,,
\end{equation}
where $P^{2\nu}_{e1}$ corresponds to the probability that an electron
neutrino produced in the core of the sun arrives at the detector as
the mass eigenstate $\nu_1$ in a two--neutrino scheme.  In contrast,
the $\mu_\mathrm{eff}^2$ relevant in reactor experiments is given as
\begin{equation} \label{finalreactor}
    \mu_\mathrm{R}^2 =
    \La^2 - \cos^2\theta_\Sol |\Lambda_1|^2 - \sin^2\theta_\Sol |\Lambda_2|^2
    -\sin 2 \theta_\Sol |\Lambda_1| |\Lambda_2| \cos\delta \,,
\end{equation}
where $\delta=\mathrm{arg}(\Lambda_1^* \Lambda_2)$ is the relative
phase between $\Lambda_1$ and $\Lambda_2$.

In the following we discuss the constraints on neutrino \TMs\ from
solar and reactor neutrino experiments~\cite{grimus:2002vb}.  The
$\chi^2$ obtained from the data is minimized with respect to all \TM\
parameters except the modulus $\La$. To take into account the physical
boundary $\La \ge 0$ we use Bayesian methods to calculate an upper
bound on $\La$. Let us stress that these bounds apply to {\it all}
elements of the \TM\ matrix, including MMs and electric dipole moments
of all neutrino flavours, since $\La^2 = |\Lambda_1|^2 +|\Lambda_2|^2
+|\Lambda_3|^2$. Furthermore, since $\La$ is independent of the basis
these bounds apply also for the \TMs\ in the flavour basis.

\begin{figure}[t]
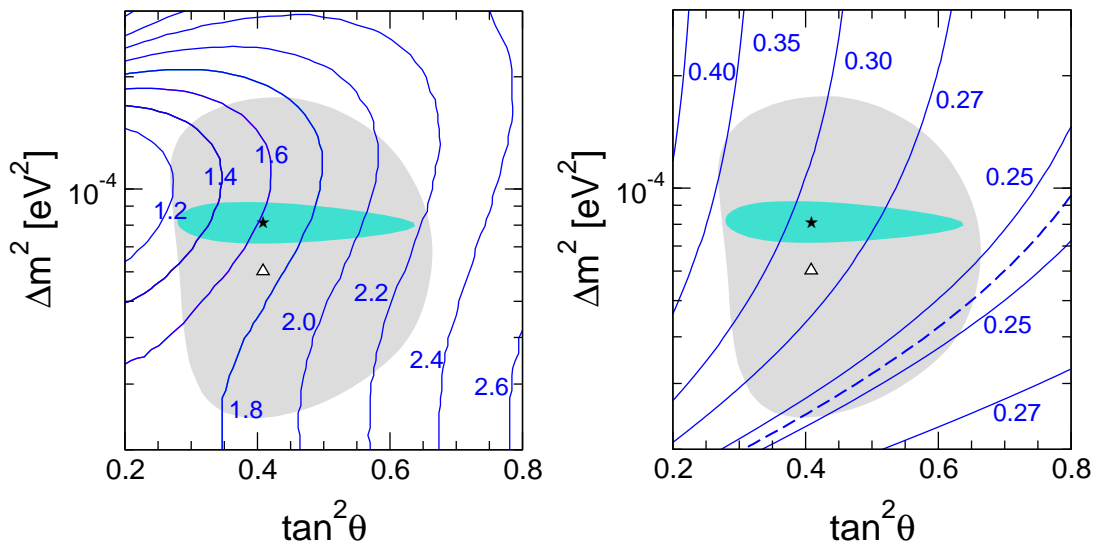
 \centering
    \includegraphics[width=0.45\linewidth]{F-cnt_glob.eps}
    \includegraphics[width=0.45\linewidth]{F-cnt_borex.eps}
    \caption{\label{fig:glob} %
      Contours of the 90\% \CL\ bound on $\La$ in units of
      $10^{-10}\mu_B$ from combined solar and reactor data (left
      panel) and after 3 years of Borexino data-taking (right panel).
      The gray (light) shaded region is the 3$\sigma$ LMA region
      obtained in the global analysis of solar neutrino data (best fit
      point marked with a triangle), whereas the green (dark) one
      corresponds to the 3$\sigma$ region obtained after including the
      KamLAND results (best fit point marked with a star). The dashed
      line in the right panel corresponds to $P_{e1}=0.5$ for $^7$Be
      neutrinos, and shows the strongest attainable limit.}
\end{figure}

In the left panel of Fig.~\ref{fig:glob} we show contours of the 90\%
\CL\ bound on $\La$ in the $(\tan^2\theta_\Sol, \Dms)$ plane for the
combination of solar and reactor data. We note that in the upper parts
of the LMA region, the solar data alone give already a strong bound on
$\La$, see Ref.~\cite{grimus:2002vb} for details.  In contrast, for
low $\Dms$ values the inclusion of reactor data plays an important
role in improving the bound. From our analysis we
find at 90\% \CL:
\begin{equation}\label{unconstrbounds}
    \La < \left\lbrace \begin{array}{l@{\quad}l}
    3.4 \times 10^{-10}\mu_B  & \mbox{(solar + KamLAND data)} \\
    1.7 \times 10^{-10}\mu_B  &
    \mbox{(solar + KamLAND + reactor data),}
\end{array}\right.
\end{equation}
where for each value of $\La$ we have minimized the $\chi^2$ with
respect to $\tan^2\theta_\Sol$ and $\Dms$.

Finally we note that the Borexino experiment~\cite{Alimonti:2000xc}
will improve these bounds roughly by one order of magnitude. This
experiment is mainly sensitive to the solar $^7$Be neutrino flux,
which will be measured by elastic neutrino--electron scattering.
Therefore, Borexino is similar to Super-K, the main difference is the
mono-energetic line of the $^7$Be neutrinos, with an energy of
0.862~MeV, which is roughly one order of magnitude smaller than the
energies of the $^8$B neutrino flux relevant in Super-K. Thanks to the
lower neutrino energy the sensitivity to electromagnetic properties is
increased, as can be seen from Eq.~\eqref{eq:cross}. Details about our
Borexino simulation can be found in Ref.~\cite{grimus:2002vb}. At the
best fit point one finds the sensitivity
\begin{equation}\label{borexinobound}
    \La \le 0.29 \times 10^{-10} \mu_B \quad\mbox{at}\quad 90\%\:\mbox{C.L.}
\end{equation}
after three years of Borexino data taking. In the right panel of
Fig.~\ref{fig:glob} we show contours of the 90\% \CL\ bound in the
$(\tan^2\theta_\Sol,\Dms)$ plane.


\section{Three--flavour neutrino oscillations}
\label{sec:three-flav-neutr}

\subsection{Global three--neutrino analysis}
\label{sec:glob-three-neutr}

In this section the three--neutrino oscillation parameters are
determined from a global analysis of the most recent neutrino
oscillation data. For earlier three--neutrino analyses see
Refs.~\cite{gonzalez-garcia:2000sq,fogli:2002au,Gonzalez-Garcia:2003qf}.
To fix the notation, we define the neutrino mass-squared differences
$\Dms \equiv \Delta m^2_{21} \equiv m^2_2 - m^2_1$ and $\Dma \equiv
\Delta m^2_{31} \equiv m^2_3 - m^2_1$, and use the convenient form of
the parameterization for the leptonic mixing matrix given in
Ref.~\cite{schechter:1980gr} and now adopted as standard by the
PDG~\cite{hagiwara:2002fs}:
\begin{equation} \label{eq:mixing}
    U=\left(
    \begin{array}{ccc}
        c_{13} c_{12}
        & s_{12} c_{13}
        & s_{13} \\
        -s_{12} c_{23} - s_{23} s_{13} c_{12}
        & c_{23} c_{12} - s_{23} s_{13} s_{12}
        & s_{23} c_{13} \\
        s_{23} s_{12} - s_{13} c_{23} c_{12}
        & -s_{23} c_{12} - s_{13} s_{12} c_{23}
        & c_{23} c_{13}
    \end{array} \right) \,,
\end{equation}
where $c_{ij} \equiv \cos\theta_{ij}$ and $s_{ij} \equiv
\sin\theta_{ij}$. Furthermore, we use the notations $\theta_{12}
\equiv \theta_\Sol$ and $\theta_{23} \equiv \theta_\Atm$. Because of
the hierarchy $\Dms \ll \Dma$ it is a good approximation to set $\Dms
= 0$ in the analysis of atmospheric and K2K data\footnote{See
  Ref.~\cite{Gonzalez-Garcia:2002mu} for a two-mass scale analysis of
  atmospheric data.}, and to set $\Dma$ to infinity for the analysis of
solar and KamLAND data. This implies furthermore that the effect of a
possible Dirac CP-violating phase~\cite{schechter:1980gr} in the
lepton mixing matrix can be neglected~\footnote{The two Majorana
  phases~\cite{schechter:1980gr} do not show up in oscillations but do
  appear in lepton number violating processes~\cite{doi:1981yb}.}.  We
perform a general fit to the global data in the five-dimensional
parameter space $s^2_{12}, s^2_{23}, s^2_{13}, \Delta m^2_{21}, \Delta
m^2_{31}$, and show projections onto various one- or two-dimensional
sub-spaces.

We include in our analysis the global solar neutrino oscillation data
from all solar neutrino experiments and the KamLAND reactor experiment
as described in Sec.~\ref{sec:dominant21}, the atmospheric neutrino
data from Super-K, as well as spectral data from the K2K long-baseline
experiment (see Sec.~\ref{sec:dominant32}). In addition we take into
account in our fit the constraints from the CHOOZ reactor
experiment~\cite{apollonio:1999ae}.

\begin{figure}[t] \centering
    \includegraphics[width=.95\linewidth]{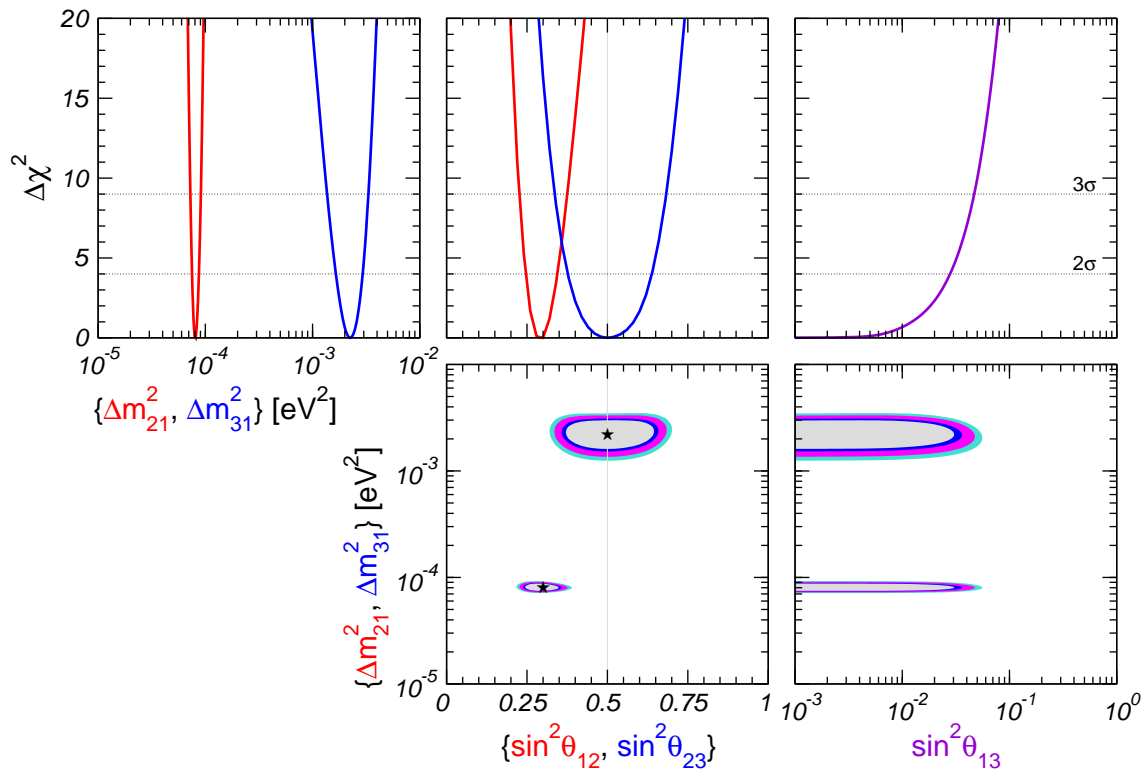}
    \caption{\label{fig:global} %
      Projections of the allowed regions from the global oscillation
      data at 90\%, 95\%, 99\%, and 3$\sigma$ \CL\ for 2 \dof\ for
      various parameter combinations. Also shown is $\Delta \chi^2$ as
      a function of the oscillation parameters $\sin^2\theta_{12},
      \sin^2\theta_{23}, \sin^2\theta_{13}, \Delta m^2_{21}, \Delta
      m^2_{31}$, minimized with respect to all undisplayed
      parameters.}
\end{figure}

\begin{table}[t] \centering
    \catcode`?=\active \def?{\hphantom{0}}
    
\begin{tabular}{|@{\quad}>{\rule[-2mm]{0pt}{6mm}}l@{\quad}|@{\quad}c@{\quad}|@{\quad}c@{\quad}|@{\quad}c@{\quad}|@{\quad}c@{\quad}|}
        \hline
        parameter & best fit & 2$\sigma$ & 3$\sigma$ & 4$\sigma$
        \\
        \hline
        $\Delta m^2_{21}\: [10^{-5}\eVq]$
        & 8.1?? & 7.5--8.7 & 7.2--9.1 & 7.0--9.4\\
        $\Delta m^2_{31}\: [10^{-3}\eVq]$
        & 2.2?? & 1.7--2.9 & 1.4--3.3 & 1.1--3.7\\
        $\sin^2\theta_{12}$
        & 0.30? & 0.25--0.34 & 0.23--0.38 & 0.21--0.41\\
        $\sin^2\theta_{23}$
        & 0.50? & 0.38--0.64 & 0.34--0.68 & 0.30--0.72 \\
        $\sin^2\theta_{13}$
        & 0.000 &  $\leq$ 0.028 & $\leq$ 0.047  & $\leq$ 0.068 \\
        \hline
    \end{tabular}
    \caption{ \label{tab:summary} Best-fit values, 2$\sigma$,
      3$\sigma$, and 4$\sigma$ intervals (1 \dof) for the
      three--flavour neutrino oscillation parameters from global data
      including solar, atmospheric, reactor (KamLAND and CHOOZ) and
      accelerator (K2K) experiments.} 
\end{table}

The results of the global three--neutrino analysis are summarized in
Fig.~\ref{fig:global} and in Tab.~\ref{tab:summary}. In the upper
panels of the figure the $\Delta \chi^2$ is shown as a function of the
parameters $\sin^2\theta_{12}, \sin^2\theta_{23}, \sin^2\theta_{13},
\Delta m^2_{21}, \Delta m^2_{31}$, minimized with respect to the
undisplayed parameters. The lower panels show two-dimensional
projections of the allowed regions in the five-dimensional parameter
space. The best fit values and the allowed ranges of the 
oscillation parameters from the global data are given in
Tab.~\ref{tab:summary}. This table summarizes the current status of
the three--flavour neutrino oscillation parameters.


\subsection{The small parameters $\alpha \equiv \Dms/\Dma$ and $\theta_{13}$}

Genuine three--flavour effects are associated to the mass hierarchy
parameter $\alpha \equiv \Dms/\Dma$ and the mixing angle
$\theta_{13}$. In particular, in a three--neutrino scheme CP violation
disappears in the limit where two neutrinos become
degenerate~\cite{schechter:1980gr,schechter:1980bn} and in the limit
where $\theta_{13} \to 0$.  We discuss in this subsection the present
status of these small parameters.

\begin{figure}[t] \centering
    \includegraphics[width=.5\linewidth]{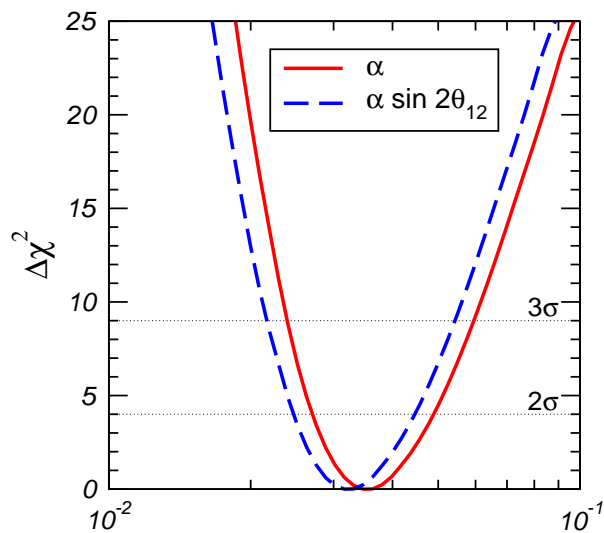}
    \caption{\label{fig:alpha}%
      $\Delta \chi^2$ from global oscillation data as a function of
      $\alpha \equiv \Dms / \Dma$ and $\alpha\sin 2\theta_{12}$.}
\end{figure}

In Fig.~\ref{fig:alpha} the $\Delta\chi^2$ from the global data is
shown as a function of the mass hierarchy parameter $\alpha$. Also
shown in this figure is the $\Delta\chi^2$ as a function of the
parameter combination $\alpha \sin 2\theta_{12}$, since to leading
order in the long baseline $\nu_e\to\nu_\mu$ oscillation probability
solar parameters appear in this particular
combination~\cite{Freund:2001pn,Akhmedov:2004ny}. We obtain the
following best fit values and 3$\sigma$ intervals:
\begin{eqnarray}
    \alpha = 0.035\,, & \quad & 0.024 \le \alpha \le 0.060\,,
    \\
    \alpha \sin 2\theta_{12} = 0.032\,, & \quad &
    0.022 \le \alpha \sin 2\theta_{12} \le 0.054\,.
    \nonumber
\end{eqnarray}

Let us now discuss the status of the mixing angle $\theta_{13}$, which
at the moment is the last unknown angle in the three--neutrino leptonic
mixing matrix. Only an upper bound exists, which used to be dominated
by the CHOOZ~\cite{apollonio:1999ae} and Palo
Verde~\cite{Boehm:2001ik} reactor experiments. Currently a large
effort is put to determine this angle in future experiments (see, \eg,
Refs.~\cite{pakvasa:2003zv,Barger:2003qi,Huber:2004ug}).

\begin{figure}[t] \centering
    \includegraphics[width=.5\linewidth]{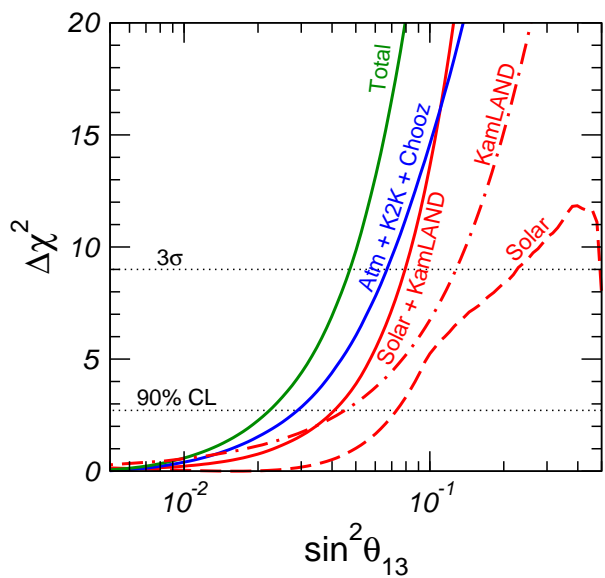}
    \caption{\label{fig:t13-compare}%
      $\Delta \chi^2$ profiles projected onto the $\sin^2\theta_{13}$
      axis, for solar, KamLAND, solar+KamLAND, atmospheric+K2K+CHOOZ, 
      and for the global data.}
\end{figure}

In Fig.~\ref{fig:t13-compare} we show the $\Delta\chi^2$ as a function
of $\sin^2\theta_{13}$ for different data sample choices. From this
figure we find the following bounds at 90\% \CL\ (3$\sigma$) for 1
\dof:
\begin{equation}\label{eq:th13}
    \sin^2\theta_{13} \le \left\lbrace \begin{array}{l@{\qquad}l}
    0.041~(0.079) & \text{(solar+KamLAND)} \\
    0.029~(0.067) & \text{(CHOOZ+atmospheric+K2K)} \\
    0.022~(0.047) & \text{(global data)}
\end{array} \right.
\end{equation}
We find that the new data from KamLAND have a surprisingly strong
impact on this bound. Before the 2004 KamLAND data the bound on
$\sin^2\theta_{13}$ from global data was dominated by the CHOOZ
reactor experiment, together with the determination of $\Delta
m^2_{31}$ from atmospheric data (see \eg, Ref.~\cite{Maltoni:2003da}).
However, using most recent data the combined bound from solar+KamLAND
becomes comparable to the CHOOZ bound, and these data contribute
notably to the final bound.  A detailed discussion of the reason for
such improvement on $\sin^2\theta_{13}$ from the 2004 KamLAND data is
given in \ref{app:discussion}. One reason is the rather strong signal
for spectral distortion in the current sample.

\begin{figure}[t] \centering
    \includegraphics[width=.5\linewidth]{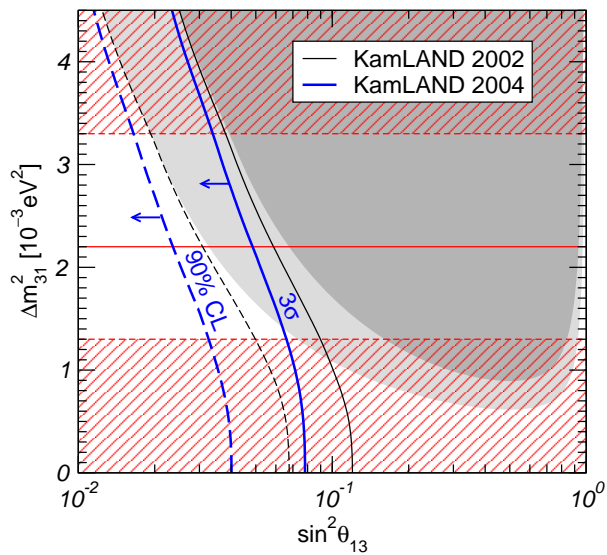}
    \caption{\label{fig:t13-solar-chooz} Upper bound on
      $\sin^2\theta_{13}$ (1 \dof) from solar+KamLAND+CHOOZ data as a
      function of $\Dma$. The dashed (solid) curve corresponds to the
      90\% (3$\sigma$) \CL\ bound, the thin curves have been obtained
      with 2002 KamLAND data, whereas the thick curves follow from the
      recent 2004 KamLAND update. The light (dark) shaded region is
      excluded by CHOOZ data alone at 90\% (3$\sigma$) \CL\ The
      horizontal line corresponds to the best fit value of $\Dma$ from
      atmospheric + K2K data given in Eq.~\eqref{eq:bfp-atm+k2k}, and
      the hatched regions are excluded by atmospheric + K2K data at
      3$\sigma$ according to Eq.~\eqref{eq:atm_ranges}.  }
\end{figure}

As noted in Ref.~\cite{Maltoni:2003da} the bound from solar+KamLAND is
especially important for the relatively lower values of $\Dma$ implied
by the use of the new three--dimensional atmospheric
fluxes~\cite{Honda:2004yz}, since the CHOOZ bound on
$\sin^2\theta_{13}$ deteriorates quickly when $\Dma$ decreases (see
Fig.~\ref{fig:t13-solar-chooz}).  Such loosening in sensitivity for
low $\Dma$ values is prevented first, by the lower bound on $\Dma$
from K2K (see Fig.~\ref{fig:k2k}) and second, by the bound from
solar+KamLAND, which is independent of $\Dma$. In
Fig.~\ref{fig:t13-solar-chooz} we show the upper bound on
$\sin^2\theta_{13}$ as a function of $\Dma$ from CHOOZ data alone
compared to the bound from an analysis including solar and reactor
neutrino data (CHOOZ and KamLAND). One finds that, although for larger
$\Dma$ values the bound on $\sin^2\theta_{13}$ is dominated by the
CHOOZ data, for $\Dma \lesssim 2 \times 10^{-3} \eVq$ the solar +
KamLAND data start being important. For illustration we show in
Fig.~\ref{fig:t13-solar-chooz} also the bound as implied by the old
2002 KamLAND data to highlight the improvement of the new data. We
note that, as before, the bound from solar data is rather stable under
the recent SSM update.

\begin{figure}[t] \centering
    \includegraphics[height=6cm,width=.95\linewidth]{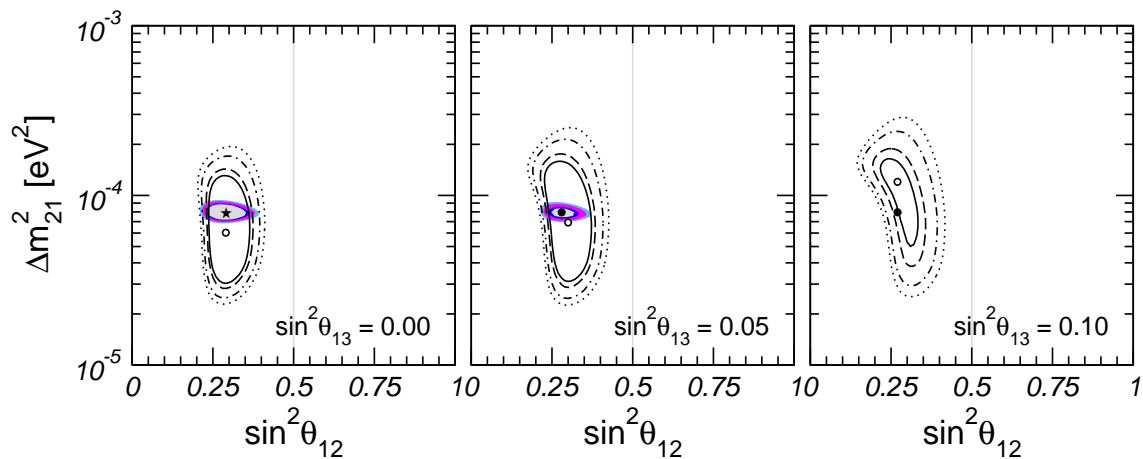}
    \caption{\label{fig:t13regions} %
      Sections of the three--dimensional allowed regions in the
      ($\sin^2\theta_\Sol,\Dms$) plane at 90\%, 95\%, 99\% and
      3$\sigma$ \CL\ for 3 \dof\ for various $\sin^2\theta_{13}$
      values from solar data (lines) and solar+KamLAND data (colored
      regions). The local minima in each plane from solar+KamLAND
      (solar only) data are marked by filled (open) dots.}
\end{figure}

Let us discuss in some more detail the constraint on
$\sin^2\theta_{13}$ from solar data, which emerges from a subtle
interplay of various solar neutrino observables. In
Fig.~\ref{fig:t13regions} we show the results of a three parameter fit
($\sin^2\theta_\Sol, \Dms$, $\sin^2\theta_{13}$) to solar and KamLAND
data. Allowed regions are shown for various values of
$\sin^2\theta_{13}$ in the ($\sin^2\theta_\Sol, \Dms$) plane with
respect to the global minimum. Note that here we calculate the allowed
regions at a given confidence level for 3 \dof\ The shape of
$\Delta\chi^2$ from solar data shown in Fig.~\ref{fig:t13-compare} can
be understood from Fig.~\ref{fig:t13regions}. Indeed one observes that
for solar data increasing $\theta_{13}$ can be compensated to some
extent by increasing $\Dms$. Since solar data disfavours large values
of $\Dms$ the bound improves. 
Also the combination with KamLAND has a similar effect, since recent
KamLAND data essentially fix $\Dms$ at roughly $8\times 10^{-5}~\eVq$
such that the continuous rise of $\Dms$ with $\sin^2\theta_{13}$
preferred by solar data is prevented. On the other hand solar data
breaks a correlation of $\sin^2\theta_{13}$ and $\sin^2\theta_{12}$ in
the KamLAND data (see \ref{app:discussion}), which again leads to an
improvement of the combined bound on $\sin^2\theta_{13}$.

The difference in the day/night solar neutrino fluxes due to the
regeneration effect in the earth in the three--flavour framework has
been considered recently in
Refs.~\cite{Akhmedov:2004rq,Blennow:2003xw}. This observable may
provide valuable information on $\theta_{13}$ in the context of future
solar neutrino experiments like UNO or Hyper-K~\cite{SKatm04}.


\setcounter{footnote}{0}

\section{Four--neutrino oscillations and LSND}
\label{sec:four-neutr-oscill}

In addition to the strong evidence for oscillations due to the
mass-squared differences $\Dms$ and $\Dma$ there is also a hint for
oscillations with a much larger mass-squared difference from the LSND
experiment~\cite{aguilar:2001ty}. This accelerator experiment
performed at Los Alamos observed $87.9\pm22.4\pm6.0$ excess events in
the $\bar\nu_\mu\to\bar\nu_e$ appearance channel, corresponding to a
transition probability of $P=(0.264\pm0.067\pm0.045)\%$, which is
$\sim 3.3\sigma$ away from zero. To explain this signal with neutrino
oscillations requires a mass-squared difference $\Dml \sim 1~\eVq$.
Such a value is inconsistent with the mass-squared differences
required by solar/KamLAND and atmospheric/K2K experiments within the
standard three--flavour framework. In this section we consider
four--neutrino schemes, where a sterile
neutrino~\cite{peltoniemi:1993ec,peltoniemi:1993ss,caldwell:1993kn} is
added to the three active ones to provide the additional mass scale
needed to reconcile the LSND evidence.
We include in our analysis data from the LSND experiment, as well as
from short-baseline (SBL)
accelerator~\cite{armbruster:2002mp,dydak:1984zq} and
reactor~\cite{Declais:1995su,apollonio:1999ae,Boehm:2001ik}
experiments reporting no evidence for oscillations (see
Ref.~\cite{grimus:2001mn} for details of our SBL data analysis). We
update our previous four--neutrino analyses (see, \eg,
Refs.~\cite{maltoni:2002xd,Schwetz:2003pv}) by including the most
recent solar and KamLAND~\cite{araki:2004mb} data, the improved
atmospheric neutrino fluxes~\cite{Honda:2004yz} and latest data from
the K2K long-baseline experiment~\cite{k2k-nu04}.

\subsection{A common parameterization for four--neutrino schemes}

\begin{figure} \centering
    \includegraphics[width=0.7\linewidth]{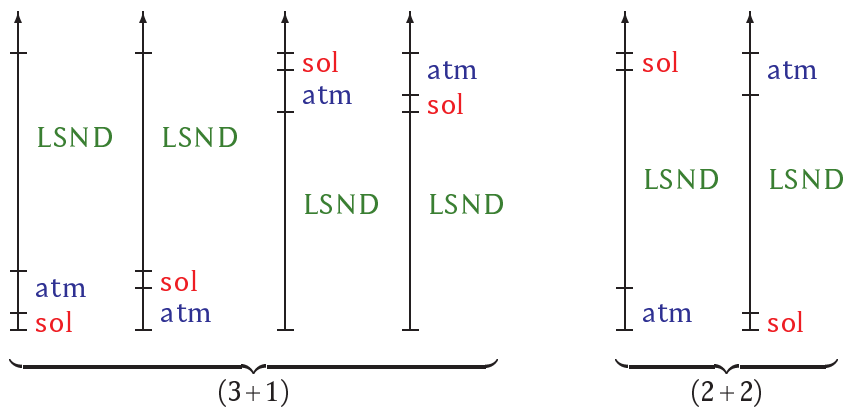}
    \caption{\label{fig:4spectra}%
      The two classes of six four--neutrino mass spectra, (3+1) and
      (2+2).}
\end{figure}

Four--neutrino mass schemes are usually divided into
the two classes (3+1) and (2+2), as illustrated in
Fig.~\ref{fig:4spectra}.  We note that (3+1) mass spectra include the
three--active neutrino scenario as limiting case. In this case solar
and atmospheric neutrino oscillations are explained by active neutrino
oscillations, with mass-squared differences $\Dms$ and $\Dma$, and the
fourth neutrino state gets completely decoupled. We will refer to such
limiting scenario as (3+0). In contrast, the (2+2) spectrum is
intrinsically different, as the sterile neutrino must take part in
either solar or in atmospheric neutrino oscillations, or in both.

Neglecting CP violation, neutrino oscillations in four--neutrino
schemes are generally described by 9 parameters: 3 mass-squared
differences and 6 mixing angles in the lepton mixing
matrix~\cite{schechter:1980gr}.  We use the parameterization
introduced in Ref.~\cite{maltoni:2001bc}, in terms of $\Dms$,
$\theta_\Sol$, $\Dma$, $\theta_\Atm$, $\Dml$, $\theta_\Lsnd$. These 6
parameters are similar to the two-neutrino mass-squared differences
and mixing angles and are directly related to the oscillations in
solar, atmospheric and the LSND experiments.  For the remaining 3
parameters we use $\eta_s,\eta_e$ and $d_\mu$. These quantities are
defined by
\begin{eqnarray}
    \label{eq:def-eta}
    \eta_\alpha &=& \sum_i |U_{\alpha i}|^2 \quad
    \mbox{with $i\in$ solar mass states,}\\%
    \label{eq:def-d}
    d_\alpha  &=& 1 - \sum_i |U_{\alpha i}|^2 \quad
    \mbox{with $i\in$ atmospheric mass states,}
\end{eqnarray}
where $\alpha = e,\mu,\tau,s$. Note that in (2+2) schemes the relation
$ \eta_\alpha = d_\alpha$ holds, whereas in (3+1) $\eta_\alpha$ and
$d_\alpha$ are independent. The physical meaning of these parameters
is the following: $\eta_\alpha$ is the fraction of $\nu_\alpha$
participating in solar oscillations, and ($1-d_\alpha$) is the
fraction of $\nu_\alpha$ participating in oscillations with $\Dma$
(for further discussions see Ref.~\cite{maltoni:2001bc}). For the
analysis we adopt the following approximations:
\begin{enumerate}
  \item
    We make use of the hierarchy $\Dms \ll \Dma \ll \Dml$.
    This means that for each data set we consider only one mass-squared
    difference, the other two are set either to zero or to infinity.
  \item In the analyses of solar and atmospheric data (but not for SBL
    data) we set $\eta_e = 1$, which is justified because of strong
    constraints from reactor
    experiments~\cite{Declais:1995su,apollonio:1999ae,Boehm:2001ik}.
\end{enumerate}

\begin{figure}[t] \centering
    \includegraphics[width=0.7\linewidth]{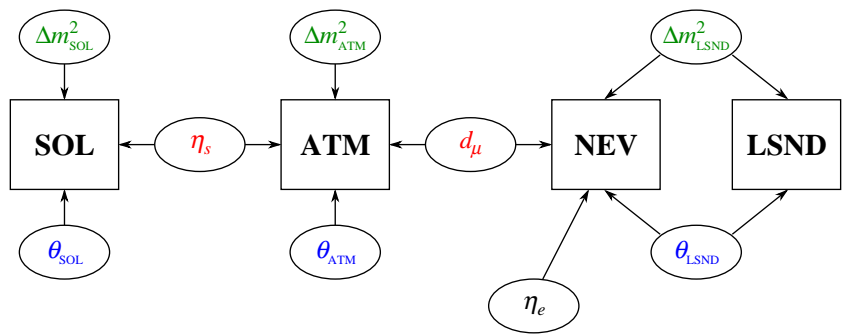}
    \caption{\label{fig:diagram}%
      Parameter dependence of the different data sets in our
      parameterization.}
\end{figure}

Within this approximation the parameter structure of the four--neutrino
analysis gets rather simple. The parameter dependence of the four data
sets solar, atmospheric, LSND and NEV is illustrated in
Fig.~\ref{fig:diagram}.  In this section, except where explicitly
noted otherwise, we tacitly consider KamLAND as part of the solar data
sample and K2K as part of the atmospheric data sample. The NEV data
set contains the experiments KARMEN~\cite{armbruster:2002mp},
CDHS~\cite{dydak:1984zq}, Bugey~\cite{Declais:1995su},
CHOOZ~\cite{apollonio:1999ae}, and Palo Verde~\cite{Boehm:2001ik},
reporting no evidence for oscillations. We see that only $\eta_s$
links solar and atmospheric data and $d_\mu$ links atmospheric and NEV
data, while LSND and NEV data are coupled by $\Dml$ and
$\theta_\Lsnd$. With the definitions \eqref{eq:def-eta} and
\eqref{eq:def-d} and in our approximation the parameter structure
shown in Fig.~\ref{fig:diagram} holds for both types of mass spectra,
(3+1) as well as (2+2)~\cite{maltoni:2001bc}.

\subsection{(2+2): ruled out by solar and atmospheric data}
\label{sec:2+2}

The strong preference for oscillations into active neutrinos in solar
and atmospheric oscillations~\cite{maltoni:2002ni} leads to a direct
conflict in (2+2) oscillation schemes. We will now show that thanks to
recent solar neutrino data (in particular from the SNO-salt
phase~\cite{Ahmed:2003kj}) in combination with the KamLAND
experiment~\cite{eguchi:2002dm}, and the latest Super-K data on
atmospheric neutrinos~\cite{fukuda:1998mi} the tension in the data has
become so strong that (2+2) oscillation schemes are essentially ruled
out~\footnote{For an earlier four--neutrino analysis of solar and
  atmospheric data see Ref.~\cite{gonzalez-garcia:2001uy}.}.

\begin{figure}[t] \centering
    \includegraphics[width=0.95\linewidth]{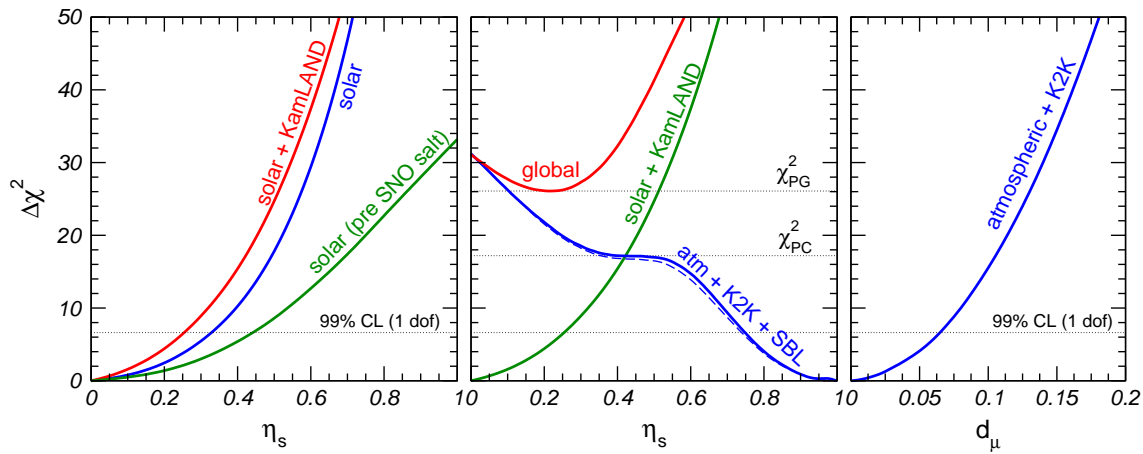}
    \caption{ \label{fig:etas} %
      Left: $\Delta\chi^2$ as a function of $\eta_s$ from solar data
      before the SNO salt-phase results, from current solar data, and
      from solar+KamLAND data. Middle: $\Delta\chi^2_\Sol$,
      $\Delta\chi^2_{\Atm+\KtK+\Sbl}$ and $\bar\chi^2_\mathrm{global}$
      as a function of $\eta_s$ in (2+2) oscillation schemes. The
      dashed line corresponds to atmospheric and K2K data only
      (without SBL data). Right: $\Delta\chi^2_{\Atm+\KtK}$ as a
      function of $d_\mu$.}
\end{figure}

In the left panel of Fig.~\ref{fig:etas} we show the $\Delta \chi^2$
from solar neutrino data as a function of $\eta_s$, the parameter
describing the fraction of the sterile neutrino participating in solar
neutrino oscillations. It is clear from the figure that the improved
determination of the neutral current event rate from the solar $^8$B
flux implied by the salt enhanced measurement in
SNO~\cite{Ahmed:2003kj} substantially tightened the constraint on a
sterile contribution: the 99\%~\CL\ bound improves from from $\eta_s
\le 0.44$ for pre-SNO-salt to $\eta_s \le 0.31$ (BP00). The boron flux
predicted in the current BP04 SSM is slightly larger than the NC flux
measured in SNO, which leaves more room for a sterile component in the
solar neutrino flux. Indeed, using the BP04 SSM the bound deteriorates
slightly to $\eta_s \le 0.33$ at the 99\%~\CL\ This effect illustrates
that in schemes beyond minimal three--flavour oscillations the data
still shows some sensitivity to the theoretical SSM input.
Although KamLAND on its own is insensitive to a sterile neutrino
contamination, it contributes indirectly to the bound because of the
better determination of $\Dms$ and
$\theta_\Sol$~\cite{maltoni:2002ni}. The combined analysis leads to
the 99\% \CL\ bound
\begin{equation}
    \eta_s \le 0.25 \qquad\mbox{(solar + KamLAND, BP04).}
\end{equation}

In contrast, in (2+2) schemes atmospheric data prefer values of
$\eta_s$ close to 1. From the combined analysis of Super-K atmospheric
data, K2K and SBL neutrino data we obtain the bound $\eta_s \ge 0.75$
at 99\%~\CL, in clear disagreement with the bound from solar data. In
the middle panel of Fig.~\ref{fig:etas} we show the $\Delta\chi^2$ for
solar data and for atmospheric+K2K combined with SBL data as a
function of $\eta_s$. Note that the main effect comes from
atmospheric+K2K data; SBL experiments contribute only marginally, as
visible from the dashed line. From this figure we also see that the
``solar+KamLAND'' and the ``atm+K2K+SBL'' allowed domains overlap only
at $\chi^2_\mathrm{PC} = 17.2$, \ie\ at the $4.1\sigma$ level.

In the middle panel of Fig.~\ref{fig:etas} we also show the ``global''
$\bar{\chi}^2$ function defined as follows:
\begin{equation}\label{eq:chi2solatm}
    \bar\chi^2(\eta_s) \equiv
    \Delta\chi^2_{\SlKm}(\eta_s) +
    \Delta\chi^2_{\Atm+\KtK+\Sbl}(\eta_s) \,.
\end{equation}
In Refs.~\cite{maltoni:2002xd,Maltoni:2003cu} we have proposed a
statistical method to evaluate the disagreement of different data sets
in global analyses. The \textit{parameter goodness of fit} (PG) makes
use of the $\bar\chi^2$ defined in Eq.~\eqref{eq:chi2solatm}. This
criterion is very useful to evaluate the GOF of the
\textit{combination} of data sets, avoiding dilution by the large
number of data points, as it happens for the usual GOF criterion (for
details see Ref.~\cite{Maltoni:2003cu}). We find $\chi^2_\mathrm{PG}
\equiv \bar\chi^2_\mathrm{min} = 26.1$, which corresponds to
5.1$\sigma$. We conclude that (2+2) mass schemes are ruled out by the
disagreement between the latest solar and atmospheric neutrino data.
This is a very robust result, independent of whether LSND is confirmed
or disproved~\footnote{Sub-leading effects beyond the approximations
adopted here should not affect this result significantly. Allowing for
additional parameters to vary at the percent level might change the
{\it ratio} of some observables~\cite{Paes:2002ah}, however, we expect
that the absolute number of events relevant for the fit will not
change substantially.}. 
  
Let us note that we now obtain a slightly smaller $\chi^2_\mathrm{PG}$
than previously~\cite{Schwetz:2003pv}, and the disagreement gets
slightly weaker. The reason is that in the present analysis we have
not included the atmospheric neutrino data from the MACRO
experiment~\cite{surdo:2002rk}. As discussed in
Ref.~\cite{maltoni:2002ni}, these data have some sensitivity to
$\eta_s$ and enhance the rejection against a sterile component in
atmospheric oscillations.
Since relevant information to perform a consistent combined analysis
of Super-K and MACRO using the new fluxes is presently not available,
we prefer to use only Super-K data, which allows us to derive simpler
and more robust results.\footnote{The reason for this is the presence
of some tension between theoretical predictions from three--dimensional
fluxes and MACRO as well as Super-K thru-going muon data (see
discussion in second reference of~\cite{surdo:2002rk}). This tension
did not appear with older one--dimensional fluxes, and makes it
particularly important to properly take into account all the
correlations between different data sets. For Super-K detailed
information on the correlation between various sub-samples (sub-GeV,
multi-GeV, stopping and thru-going muons) can be extracted from
Ref.~\cite{kameda}, whereas the relevant information to include MACRO
is not available. Since in this analysis we are more interested in the
\emph{robustness} rather than in the \emph{strength} of our results,
we restrict our analysis to Super-K data only. A consistent way of
combining Super-K and MACRO data using three--dimensional fluxes is
however under investigation.} Let us mention that we are also
neglecting neutral current and tau appearance data from Super-K,
which would also increase the rejection against a sterile component,
since the detailed informations which are needed to use these data are
presently not available outside the Super-K collaboration. 

\subsection{(3+1): strongly disfavoured by SBL data}
\label{sec:3+1}

\begin{figure}[t] \centering
    \includegraphics[height=8cm]{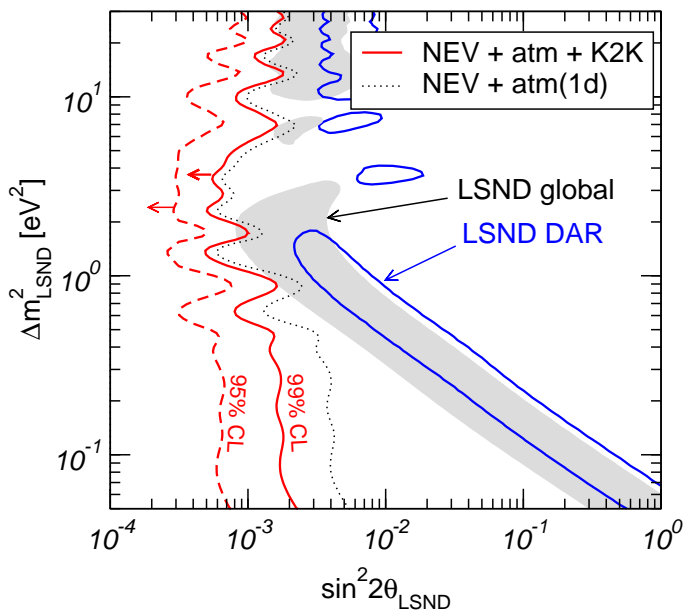}
    \caption{\label{fig:3+1} %
      Upper bound on $\sin^22\theta_\Lsnd$ from NEV, atmospheric and
      K2K neutrino data in (3+1) schemes. The bound is calculated for
      each $\Dml$ using the $\Delta \chi^2$ for 1~\dof\ The dotted
      line corresponds to the bound at 99\%~\CL\ without K2K and using
      one--dimensional atmospheric fluxes. Also shown are the regions
      allowed at 99\% \CL\ (2~\dof) from global LSND
      data~\protect\cite{aguilar:2001ty} and decay-at-rest (DAR) LSND
      data~\protect\cite{Church:2002tc}.}
\end{figure}

It is known for a long
time~\cite{bilenky:1998rw,okada:1997kw,barger:1998bn,bilenky:1999ny,peres:2000ic,giunti:2000ur,grimus:2001mn}
that (3+1) mass schemes are disfavoured by the comparison of SBL
disappearance data~\cite{dydak:1984zq,Declais:1995su} with the LSND
result. The reason is that in (3+1) schemes the relation
$\sin^22\theta_\Lsnd = 4\, d_e\,d_\mu$ holds, and the parameters $d_e$
and $d_\mu$ (see Eq.~(\ref{eq:def-d})) are strongly constrained by
$\nu_e$ and $\nu_\mu$ disappearance experiments, leading to a double
suppression of the LSND amplitude. In Ref.~\cite{bilenky:1999ny} it
was realized that the up-down asymmetry observed in atmospheric $\mu$
events leads to an additional constraint on $d_\mu$. The
$\Delta\chi^2(d_\mu)$ from the fit to atmospheric+K2K data is shown in
the right panel of Fig.~\ref{fig:etas}. We find that the use of the
new atmospheric fluxes~\cite{Honda:2004yz} as well as K2K
data~\cite{k2k-nu04} considerably strengthen the constraint on
$d_\mu$: the new bound $d_\mu \le 0.065$ at 99\%~\CL\ decreases
roughly a factor 2 with respect to the previous bound $d_\mu \le
0.13$~\cite{maltoni:2001mt,maltoni:2002ni}, as implied by fitting
atmospheric data with one--dimensional fluxes~\cite{barr:1989ru} and
without K2K. Following Ref.~\cite{maltoni:2001mt} we show in
Fig.~\ref{fig:3+1} the upper bound on the LSND oscillation amplitude
$\sin^22\theta_\Lsnd$ from the combined analysis of NEV and
atmospheric neutrino data. This figure illustrates that the
improvement implied by the stronger bound on $d_\mu$ is mostly
relevant for lower values of $\Dml$. From this figure we see that the
bound is incompatible with the signal observed in LSND at the
95\%~\CL\ Only marginal overlap regions exist between the bound and
global LSND data if both are taken at 99\%~\CL\ Using only the
decay-at-rest LSND data sample~\cite{Church:2002tc} the disagreement
is even more severe. These results show that (3+1) schemes are
strongly disfavoured by SBL disappearance data.

\subsection{Comparing (3+1), (2+2) and (3+0) hypotheses}

Using the methods developed in Ref.~\cite{maltoni:2001bc} we perform a
global fit to the oscillation data in the four--neutrino framework.
This approach allows to statistically compare the different
hypotheses. Let us first evaluate the GOF of (3+1) and (2+2) spectra
using the PG method described in Ref.~\cite{Maltoni:2003cu}. We divide
the global oscillation data into the four data sets SOL, ATM, LSND and
NEV. Then we use the PG method to test the statistical compatibility
of these data sets assuming a given neutrino mass scheme. Following
Ref.~\cite{maltoni:2002xd} we consider
\begin{equation}\label{eq:chi2bar}
    \begin{array}{ccl}
        \bar\chi^2 &=&
        \Delta\chi^2_\Sol(\theta_\Sol,\Dms,\eta_s)
        + \Delta\chi^2_\Atm(\theta_\Atm,\Dma,\eta_s,d_\mu) \\
        &+& \Delta\chi^2_\Nev(\theta_\Lsnd,\Dml,d_\mu,\eta_e)
        + \Delta\chi^2_\Lsnd(\theta_\Lsnd,\Dml) \,,
    \end{array}
\end{equation}
where $\Delta\chi^2_X = \chi^2_X - (\chi^2_X)_\mathrm{min}$ ($X$ =
SOL, ATM, NEV, LSND). In Tab.~\ref{tab:pg} we show the contributions
of the 4 data sets to $\chi^2_\mathrm{PG} \equiv
\bar\chi^2_\mathrm{min}$ for (3+1) and (2+2) oscillation schemes. As
expected we observe that in (3+1) schemes the main contribution comes
from SBL data due to the tension between LSND and NEV data in these
schemes.  For (2+2) oscillation schemes a large part of
$\chi^2_\mathrm{PG}$ comes from solar and atmospheric data, due to the
rejection against a sterile neutrino contribution of these two data
sets, as discussed in Sec.~\ref{sec:2+2}. The contribution from NEV
data in (2+2) comes mainly from the tension between LSND and
KARMEN~\cite{Church:2002tc}, which does not depend on the mass scheme.

\begin{table}[t]\centering
    \catcode`?=\active \def?{\hphantom{0}}
    \begin{tabular}{|c|cccc|c|c|}
        \hline
        & SOL & ATM & LSND & NEV &   $\chi^2_\mathrm{PG}$ & PG \\
        \hline
        (3+1) & 0.0 & ?0.4 & 5.7 & 10.9 & 17.0 & $1.9 \times 10^{-3} \: (3.1\sigma)$ \\
        (2+2) & 5.3 & 20.8 & 0.6 & ?7.3 & 33.9 & $7.8 \times 10^{-7} \: (4.9\sigma)$ \\
        \hline
    \end{tabular}
    \caption{Parameter GOF and the contributions of different data sets to
      $\chi^2_\mathrm{PG}$ in (3+1) and (2+2) neutrino mass schemes.}
    \label{tab:pg}
\end{table}

The parameter goodness of fit is obtained by evaluating
$\chi^2_\mathrm{PG}$ for 4 \dof~\cite{Maltoni:2003cu}. This number of
degrees of freedom corresponds to the 4 parameters $\eta_s, d_\mu,
\theta_\Lsnd, \Dml$ describing the coupling of the different data sets
(see Eq.~\eqref{eq:chi2bar} and Fig.~\ref{fig:diagram}). The best GOF
is obtained in the (3+1) case. However, even in this best case the PG
is only 0.19\%. This PG value is slightly reduced with respect to our
previous result 0.56\%~\cite{Schwetz:2003pv,maltoni:2002xd} because of
the improved limit on $d_\mu$ from atmospheric+K2K data (see
Sec.~\ref{sec:3+1}). The PG of $7.8 \times 10^{-7}$ for (2+2) schemes
shows that these mass schemes are essentially ruled out by the
disagreement between the individual data sets.  As mentioned in
Sec.~\ref{sec:2+2} the exclusion of (2+2) schemes is slightly weaker
as previously~\cite{Schwetz:2003pv} since we now do not include MACRO
data in our analysis.

Although we have seen that none of the four--neutrino mass schemes
provides a good fit to the global oscillation data including LSND, it
is interesting to consider the \textit{relative} status of the three
hypotheses (3+1), (2+2) and the three--active neutrino scenario (3+0).
This can be done by comparing the $\chi^2$ value of the best fit point
-- which occurs for the (3+1) scheme -- with the ones corresponding to
(2+2) and (3+0). First we observe that (2+2) schemes are strongly
disfavoured with respect to (3+1) with a $\Delta \chi^2 = 16.9$. For 4
\dof\ this is equivalent to an exclusion at $3.1\sigma$.  Furthermore,
we find that (3+0) is disfavoured with a $\Delta \chi^2 = 17.5$
(corresponding to $3.2\sigma$ for 4 \dof) with respect to (3+1).  This
reflects the high statistical significance of the LSND result, since
in a (3+0) scheme no effect is predicted for LSND.

To summarize, we find that four--neutrino schemes do not provide a
satisfactory fit to the global data.
The strong rejection of non-active oscillation in the solar+KamLAND
and atmospheric+K2K neutrino data rules out (2+2) schemes,
irrespective of whether LSND is confirmed or not. Using an improved
goodness of fit method especially sensitive to the combination of data
sets we find that (2+2) schemes are ruled out at the $4.9\sigma$
level.
On the other hand (3+1) spectra are disfavoured by the disagreement of
LSND with short-baseline disappearance data, leading to a marginal GOF
of $1.9\times 10^{-3}$ ($3.1\sigma$). Should LSND be confirmed it
would be very desirable to have more data on $\nu_e$ and/or $\nu_\mu$
SBL disappearance to decide about the status of (3+1) schemes. In that
case a positive signal is predicted right at the sensitivity edge of
existing experiments.

More drastic attempts to reconcile the LSND signal with the rest of
neutrino oscillation data have been reviewed in
Ref.~\cite{pakvasa:2003zv}.  For example, in Ref.~\cite{Sorel:2003hf}
a five--neutrino scheme is invoked to reconcile all the data. In
Ref.~\cite{Gonzalez-Garcia:2003jq} it has been shown that even the
rather drastic assumption of CPT violation in a three--neutrino
framework~\cite{murayama:2000hm,Barenboim:2002ah} does not provide a
satisfactory description of the global neutrino data set including
LSND.  Similarly, an interpretation of the LSND signal in terms of a
non-standard muon decay is disfavoured by
KARMEN~\cite{Armbruster:2003pq}.

We conclude that currently no convincing explanation for the LSND
result exists, and it remains a puzzle how to reconcile this evidence
with the rest of the data. It is therefore very important to settle
this issue experimentally. A confirmation of the LSND signal by the
MiniBooNE experiment~\cite{zimmerman:2002xj} would be very exciting
and would require some novel physics ideas.

\section{Summary and conclusions}
\label{sec:summary-conclusions}

We have given a brief review of the status of global analyses of
neutrino oscillations, taking into account the latest neutrino data,
including the most recent updates of KamLAND and K2K presented at
Neutrino2004, as well as state-of-the-art solar and atmospheric
neutrino flux calculations.  We presented two-neutrino solar + KamLAND
results, as well as two-neutrino atmospheric + K2K oscillation
regions, and a discussion in each case of the robustness with which
the oscillation hypothesis can be established, in view of possible
modifications. These might come from the assumed theoretical fluxes,
the non-validity of the Standard Model neutrino interaction cross
sections or the existence of non-trivial neutrino propagation
properties beyond oscillations. As case studies we have mentioned the
robustness of the solar neutrino oscillation hypothesis \textsl{vis a
vis} the possible existence of radiative-zone solar density
fluctuations, nonzero convective--zone solar magnetic fields and
neutrino transition magnetic moments.  For the atmospheric + K2K
analysis we have considered explicitly the robustness of the
oscillation hypothesis against the possible existence of flavour or
universality violating non-standard neutrino interactions.

Furthermore, we have performed a fit to the most recent world neutrino
data sample in the three--flavour framework. The results of this global
analysis are summarized in Fig.~\ref{fig:global} and
Tab.~\ref{tab:summary}, where we give the best fit values and allowed
ranges of the three--flavour oscillation parameters. In addition we
discussed in detail the status of the small parameters $\alpha \equiv
\Dms/\Dma$ and $\sin^2\theta_{13}$, which characterize the strength of
CP violating effects in neutrino oscillations, highlighting the
improvement of the bound on $\sin^2\theta_{13}$ implied by the
inclusion of the recent KamLAND data. Finally, we gave a review over
the current status of four--neutrino interpretations of the LSND
anomaly, in view of the most recent experimental and theoretical
advances.

All in all, we can say beyond reasonable doubt that neutrino masses,
discovered through atmospheric neutrino oscillations, have now also
been confirmed in the solar neutrino oscillation channel thanks to the
important input of the KamLAND experiment.  Theory-wise, while the SSM
was necessary in order to establish the need for physics beyond the
Standard Model, it has now been made to some extent irrelevant by the
high precision of the experiments which currently dominate the
determination of solar neutrino oscillation parameters. The next goal
in the agenda is the determination of the small parameter
$\sin^2\theta_{13}$ that characterizes the strength of CP violating
effects in neutrino oscillations, and the exploration of the Majorana
nature of the neutrino which will be sensitive to the other leptonic
CP phases.

\section*{Acknowledgments}

This review is based on work with a number of collaborators with whom
we have had many (sometimes passionate) discussions. The list includes
E.Kh. Akhmedov, J. Barranco, C.P. Burgess, N.S. Dzhalilov, N.
Fornengo, M.C. Gonz\'alez-Garc{\'\i}a, W. Grimus, M. Guzzo, P.C. de
Holanda, O. Miranda, H. Nunokawa, C. Pe{\~n}a-Garay, T.I. Rashba, A.I.
Rez, V.B. Semikoz and R. Tom\`as. 
This work was supported by Spanish grant BFM2002-00345,
by the European Commission RTN grant HPRN-CT-2000-00148 and the ESF
\emph{Neutrino Astrophysics Network}. M.M.\ is supported by the
National Science Foundation grant PHY0098527. T.S.\ has been supported
by the ``Sonderforschungsbereich 375 f{\"u}r Astro-Teilchenphysik der
Deutschen For\-schungs\-ge\-mein\-schaft''. M.T.\ has been supported
by FPU fellowship AP2000-1953.

\begin{appendix}

\section{The new KamLAND results}
\label{app:kamland}

In this appendix we discuss in some detail the most recent data from
KamLAND~\cite{araki:2004mb}. In \ref{app:details} we give the details
of our data analysis, whereas in \ref{app:discussion} we discuss the
physics results.

\subsection{Analysis details of the new data}
\label{app:details}

We have modified our previous KamLAND analysis
methods~\cite{maltoni:2002aw,Schwetz:2003se} applied to the first
published KamLAND data~\cite{eguchi:2002dm} in various aspects to take
into account the characteristics of the new data.
First, to analyze KamLAND data one has to know the contribution of the
various power reactors to the signal. We extract the relevant
information from Fig.~1(b) of Ref.~\cite{araki:2004mb}, where the
no-oscillation signal is given as a function of the distance to the
detector. Second, we use an improved
parameterization~\cite{Huber:2004xh} of the anti-neutrino flux emitted
by the isotopes $^{235}$U, $^{239}$Pu, $^{238}$U and $^{241}$Pu in the
nuclear reactors. Third, we include 2.69 accidental background events
in the lowest energy bin. For the $4.8\pm0.9$ background events
expected from the beta-decay of $^9$Li and $^8$He we assume regular
beta-spectra with endpoints of 13.6 and 10.7~MeV, respectively (see,
\eg, Fig.~11 of Ref.~\cite{Huber:2004ug}). Fourth, we include the small
matter effects.

\begin{figure}[t] \centering
    \includegraphics[height=9cm]{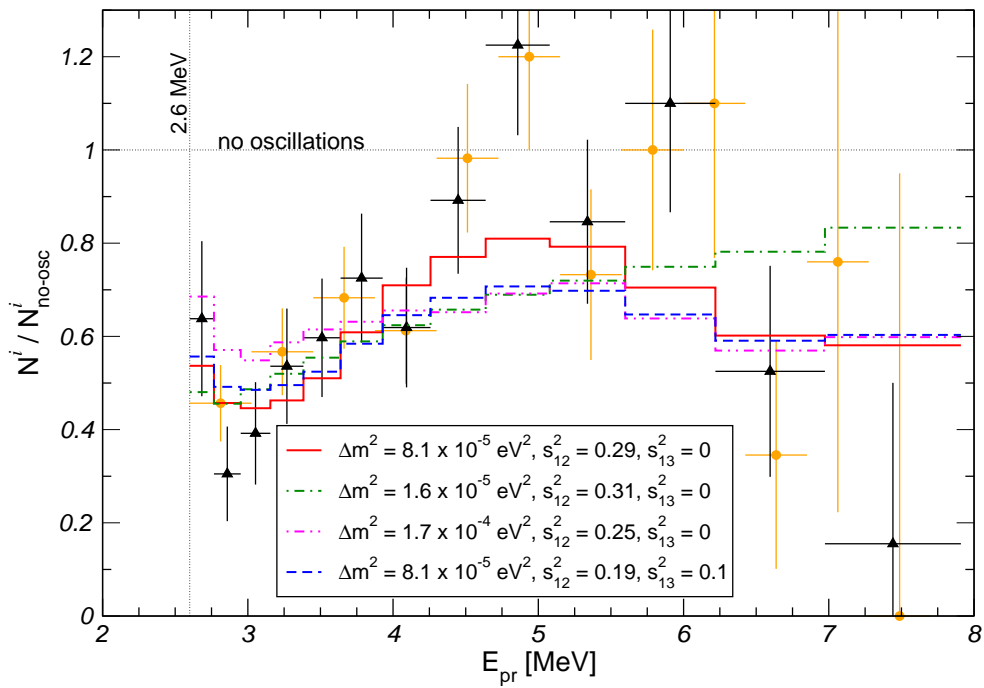}
    \caption{\label{fig:KL-spectrum} Data divided by the expectation for
      no oscillations for equal bins in $1/E_\mathrm{pr}$ (triangles)
      and in $E_\mathrm{pr}$ (dots). Also shown is the spectrum for
      the best fit point (solid), the local best fit points in the low
      $\Delta m^2$ (dash-dotted) and high $\Delta m^2$
      (dash-dot-dotted) regions, and for a large value of
      $\sin^2\theta_{13}$ (dashed).}
\end{figure}

In general most information can be extracted from data by un-binned
Likelihood methods (see Ref.~\cite{Schwetz:2003se} for the case of the
first KamLAND data). Unfortunately it is not possible to obtain
event-based energy information for the current KamLAND sample, and one
has to stick with binned data outside the collaboration. Traditionally
data is given in bins of equal size in the prompt energy
$E_\mathrm{pr}$ (see Fig.~2(b) of Ref.~\cite{araki:2004mb}). However,
it turns out that in the case of KamLAND more information can be
obtained if data is binned equally in $1/E_\mathrm{pr}$. The relevant
information can be extracted from Fig.~3 of Ref.~\cite{araki:2004mb},
where the ratio of the observed spectrum to the expectation for no
oscillation is given in 13 bins of $180\,\mathrm{km}/
E_\nu\,\mathrm{[MeV]}$. We show the data binned in $1/E_\mathrm{pr}$
as well as in $E_\mathrm{pr}$ in Fig.~\ref{fig:KL-spectrum}. The
$1/E_\mathrm{pr}$ binning is more useful for two reasons. First, it is
more natural to make the bins smaller in the region of many events
(low energy) and wider in the high energy region, where there are very
few events. This maintains more energy information in the region of
greater statistics. Second, since the frequency of neutrino
oscillations is proportional to $1/E_\nu$ this binning is more
appropriate for the signal we are interested.

Concerning the statistical analysis, we adopt a Poisson
$\chi^2$-function, and make extensive use of the pull-method to
implement various systematical errors. In the overall normalization
uncertainty we include only the detector--specific contributions by
summing up the errors of the left column of Tab.~1 from
Ref.~\cite{araki:2004mb}, which gives $\sigma_\mathrm{det} = 5.47\%$.
The uncertainties associated to the anti-neutrino flux are treated
according to the method presented in Ref.~\cite{Huber:2004xh}. We
include an uncertainty on the thermal power (2\%) and the fuel
composition (1\%) of each individual reactor, as well as the spectral
uncertainty of the emitted anti-neutrino fluxes. In agreement with
Ref.~\cite{Huber:2004xh} we find that flux related uncertainties play
only a minor role in the KamLAND analysis. 

\subsection{Discussion of the KamLAND results}
\label{app:discussion}

In Fig.~\ref{fig:KL-spectrum} we show the predictions for the
probabilities in the various bins $P_i \equiv N^i_\mathrm{osc} /
N^i_\mathrm{no-osc}$ compared to the data $P^\mathrm{obs}_i \equiv
N^i_\mathrm{obs} / N^i_\mathrm{no-osc}$. From this figure one can see
that the probabilities are rather low for low energies, whereas in
the region around 4.5~MeV even slightly more events than expected for
no oscillations have been observed. This is a very characteristic
pattern indicating rather strong spectral distortion, and only the
best fit parameters (solid line) can reproduce this shape. All other
parameter choices shown in the figure have problems to accommodate the
high data points in the middle of the spectrum. This holds for the
solution around $\Delta m^2 \simeq 1.6\times 10^{-5}$~eV$^2$ as well
as for the high-LMA solution at $\Delta m^2 \simeq 1.7\times
10^{-4}$~eV$^2$ (compare Fig.~\ref{fig:kaml-region}). The rejection
power to these two ``solutions'' is significantly increased by the
$1/E_\mathrm{pr}$ binning with respect to the $E_\mathrm{pr}$ binning:
In our analysis we obtain a $\Delta \chi^2 = 7.5 (11.3)$ for the local
minimum at $1.6 (17) \times 10^{-5}$~eV$^2$, whereas with the
traditional $E_\mathrm{pr}$ binning we find $\Delta \chi^2 = 4.0
(8.5)$. In particular, to disfavour the high-LMA region the more
precise energy information at low energies provided by the
$1/E_\mathrm{pr}$ bins is crucial, whereas for the low region the high
energy part seems to be important.

\begin{figure}[t] \centering
    \includegraphics[width=\textwidth]{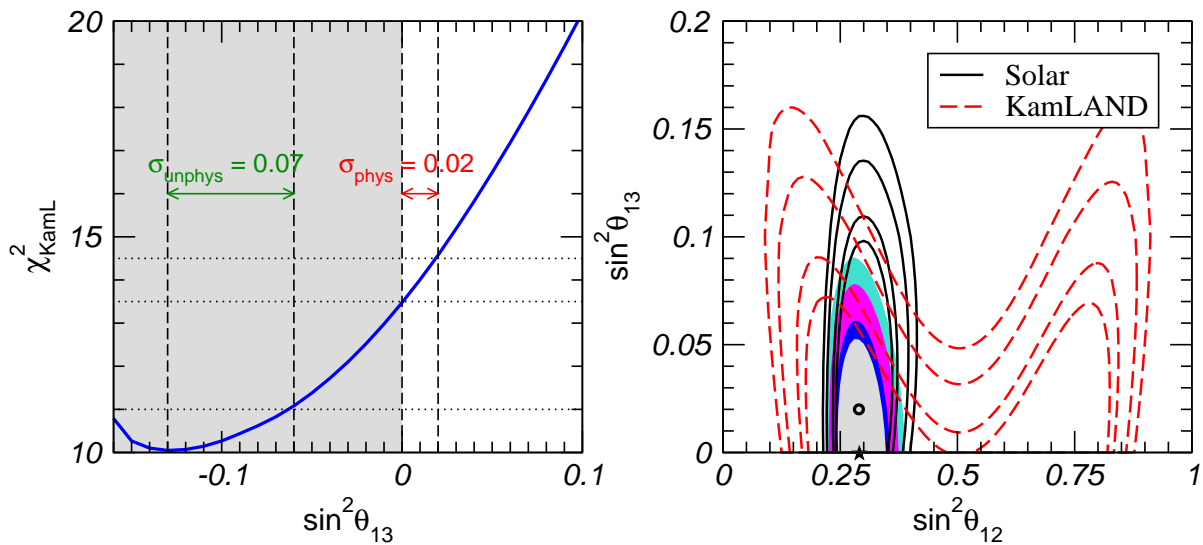}
    \caption{\label{fig:KL-th13} Left panel: $\chi^2$ from KamLAND
    data as a function of $\sin^2\theta_{13}$ including the unphysical
    region $\sin^2\theta_{13} < 0$. Right panel: allowed regions at
    90\%, 95\%, 99\% and 3$\sigma$ of $\sin^2\theta_{13}$ and
    $\sin^2\theta_{12}$ (2 \dof) for fixed $\Dms = 8.1 \times
    10^{-5}~\eVq$ from KamLAND data (dashed), solar data (solid), and
    KamLAND+solar data combined (shaded regions).}
\end{figure}

Let us now discuss the bound on $\sin^2\theta_{13}$ from the KamLAND
data.  From Fig.~\ref{fig:KL-spectrum} one can see that turning on
$\sin^2\theta_{13}$ leads to a flatter energy spectrum and it gets
more difficult to accommodate the low probabilities for low energies
simultaneously with the high data points in the middle part of the
spectrum. It is also clear from the relevant three--flavour survival
probability in vacuum
\begin{equation}\label{eq:KLprob}
P_{ee} = 1 - \frac{1}{2} \sin^2 2 \theta_{13} - 
\cos^4 \theta_{13} \,
\sin^2 2 \theta_{12} \, 
\sin^2\frac{\Delta m^2_{21} L}{4E_\nu}
\end{equation}
that values of $\sin^2\theta_{13} > 0$ suppress the oscillatory term.
In fact, if a fit to the KamLAND data is performed without imposing
the constraint $\sin^2\theta_{13} \ge 0$ one finds a best fit point
within the unphysical region at $\sin^2\theta_{13} = - 0.13$ (see
Fig.~\ref{fig:KL-th13}). Because of the rather large difference in the
probability between the low and medium energy bins the fit is improved
by a $\Delta \chi^2 \simeq 3.5$ by allowing an enhancement of the
oscillatory term in Eq.~(\ref{eq:KLprob}) due to values of
$\cos^4\theta_{13} > 1$, \ie, $\sin^2\theta_{13} < 0$.
Fig.~\ref{fig:KL-th13} illustrates that this fact leads to a rather
strong bound on $\sin^2\theta_{13}$ if the analysis is restricted to
the physical region. Just from the statistical power of the data one
would expect a $1\sigma$ error on $\sin^2\theta_{13}$ of $\sigma
\simeq 0.07$. However, due to the particular fluctuation observed in
the actual data the best fit point lies in the unphysical region. This
implies that for $\sin^2\theta_{13} = 0$ there is already some tension
in the fit, and a significantly smaller error of $\sigma = 0.02$ is
obtained within the physical region.\footnote{We note that in such a
case a reliable bound can be calculated by performing a Monte Carlo
simulation of many synthetic data sets. This is beyond the scope of
the present article, and we throughout use the naive method of
calculating bounds by considering $\Delta\chi^2$-values as implied by
Gaussian statistics and restricting the analysis to the physical
region.}

Finally, in the right panel of Fig.~\ref{fig:KL-th13} we illustrate,
why the combination of KamLAND and solar data leads to a further
notable improvement of the bound on $\sin^2\theta_{13}$. From
Eq.~(\ref{eq:KLprob}) one expects for KamLAND a correlation between
$\sin^2\theta_{13}$ and $\sin^2 2\theta_{12}$. 
Solving Eq.~(\ref{eq:KLprob}) for $\sin^2
2\theta_{12}$ and expanding up to first order in $\sin^2\theta_{13}$ one
finds
\begin{equation}\label{eq:sq13}
\sin^2 2 \theta_{12} \,\sin^2\frac{\Delta m^2_{21} L}{4E_\nu}
\simeq
1 - P_{ee} - 2 P_{ee} \sin^2\theta_{13}
\end{equation}
For a given survival probability Eq.~(\ref{eq:sq13}) implies a negative
correlation between $\sin^2\theta_{13}$ and $\sin^2 2\theta_{12}$.  This
trend is visible also in the actual fit, see Fig.~\ref{fig:KL-th13}, where
the allowed region for these two parameters is shown for fixed $\Delta
m^2$. One observes that increasing $\sin^2\theta_{13}$ can be
compensated to some extent by decreasing $\sin^2\theta_{12}$ (within
the ``light side'' $\sin^2\theta_{12} < 0.5$). This, however, is in
disagreement with solar data, which provide a stable lower bound on
$\sin^2\theta_{12}$ due to the fundamentally different conversion
mechanism (MSW matter effect in the sun versus vacuum
oscillations). The combined analysis of KamLAND and solar data leads
essentially to the intersection of the two individual allowed regions,
which further improves the bound on $\sin^2\theta_{13}$.

\section{Implications of a new background in KamLAND}

After this paper has been published it was realised that a background from the
reaction $^{13}\mathrm{C}(\alpha,n)^{16}\mathrm{O}$ contributes to the KamLAND
data, which has not been taken into account in the first version of
Ref.~\cite{araki:2004mb}, on which our analysis is based. In this appendix we
show that the impact of this new background on our results is small, and we
present the updated results, where the changes are noticeable.

\begin{figure}[t] \centering
    \includegraphics[width=\textwidth]{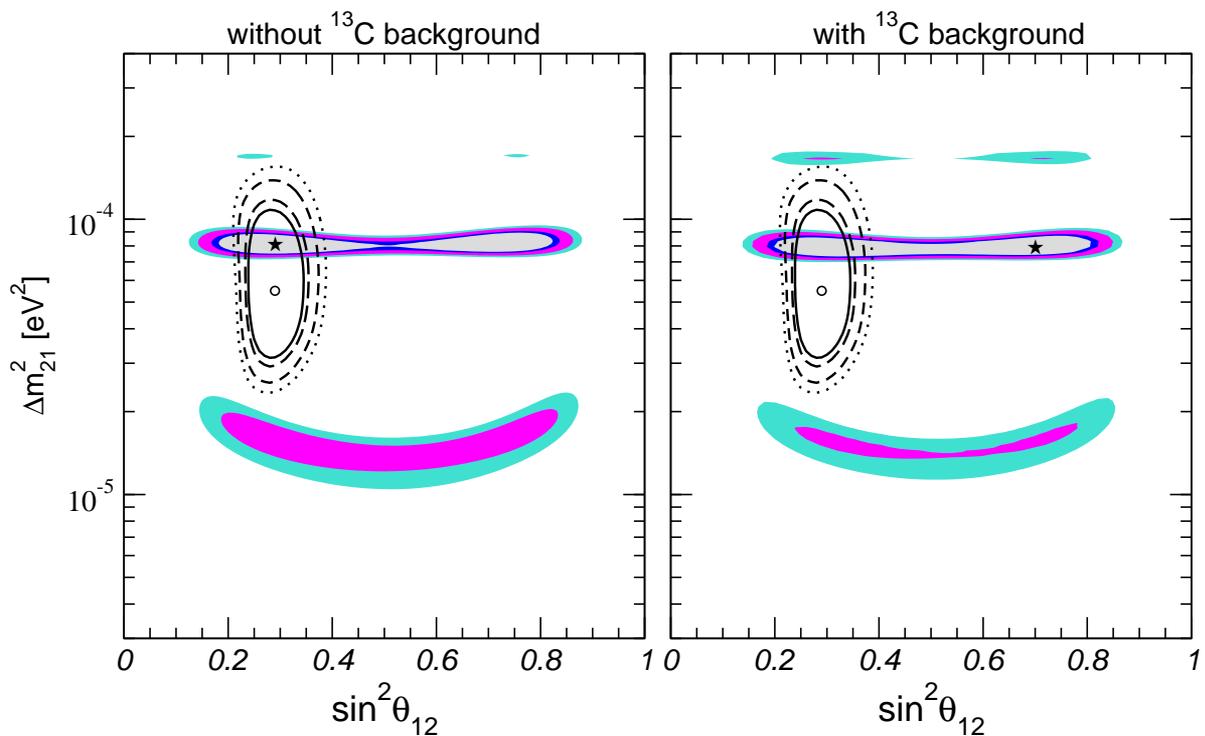}
    \caption{\label{fig:C-BG-solarparams} Allowed
      ($\sin^2\theta_\Sol$,~$\Dms$) regions at 90\%, 95\%, 99\%, and 3$\sigma$
      \CL\ for 2 \dof\ The left panel corresponds to
      Fig.~\ref{fig:kaml-region}, whereas in the right panel the new
      background from $^{13}$C has been taken into account.}
\end{figure}

The reaction $^{13}\mathrm{C}(\alpha,n)^{16}\mathrm{O}$ leads to $10.3\pm7.1$
events above the 2.6~MeV threshold in KamLAND, and hence the total background
is increased to $17.8\pm7.3$ events. This new background is mainly
concentrated around 6~MeV, and the main effect of subtracting these events
from the reactor data is that the relatively high data point at 6~MeV visible
in Fig.~\ref{fig:KL-spectrum} is moved from a value of 1.1 to 0.8, in better
agreement with the prediction for oscillation. Indeed, we now find
$\chi^2_\mathrm{min} = 9.5$ instead of $13.5$, \ie\ the quality of the fit
improves. The impact of the new background on the determination of the solar
parameters is shown in Fig.~\ref{fig:C-BG-solarparams}. The best fit point for
$\Dms$ moves from $8.1\times 10^{-5}$~eV$^2$ to $7.9\times
10^{-5}$~eV$^2$. However, the allowed region around the best fit point is very
stable. The fact that the best fit point occurs now for $\sin^2\theta_\Sol >
0.5$ has no statistical significance. The low-$\Dms$ solution is slightly more
disfavoured, with $\Delta\chi^2 = 8.6$ instead of $7.4$, whereas the the
high-$\Dms$ solution becomes somewhat better, with $\Delta\chi^2 = 8.8$
instead of $11.2$. Since these solutions are ruled out anyway by solar data,
the overall changes in the global fit are very small.

\begin{figure}[t]
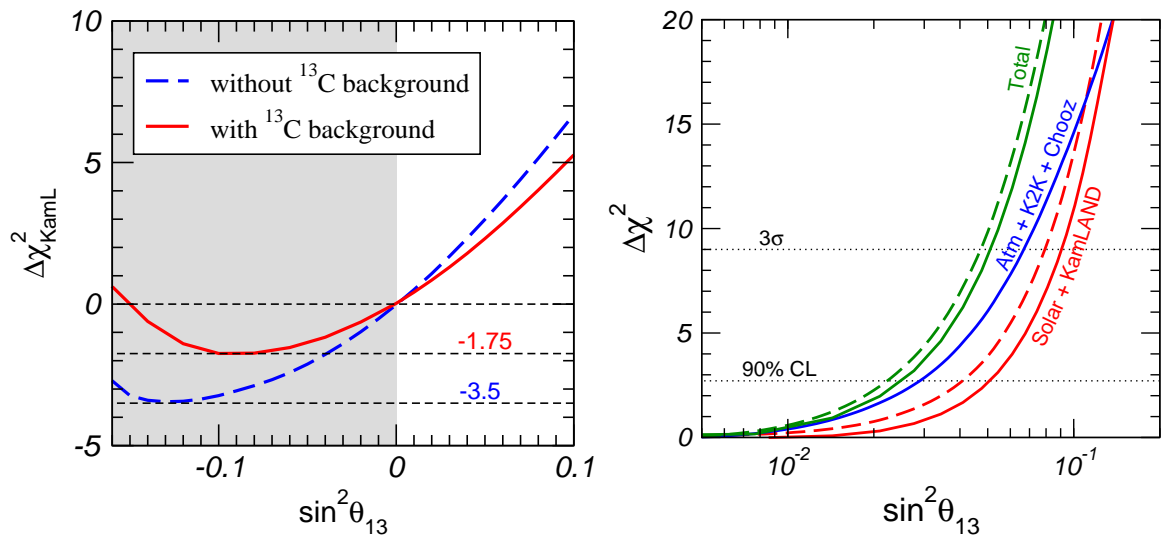
 \centering
    \includegraphics[height=7cm]{th13-unphys.eps}
    \includegraphics[height=7cm]{F-th13-chisq-C-BG.eps}
    \caption{\label{fig:C-BG-th13} Left panel: $\Delta\chi^2$ from KamLAND
    data allowing for negative values of $\sin^2\theta_{13}$ with and without
    accounting for the $^{13}$C background. Right panel: $\Delta\chi^2$ as a
    function of $\sin^2\theta_{13}$ for various data samples. The $^{13}$C
    background is (is not) taken into account for the solid (dashed) curves.}
\end{figure}

\begin{table}[t] \centering
    \catcode`?=\active \def?{\hphantom{0}}
    
\begin{tabular}{|@{\quad}>{\rule[-2mm]{0pt}{6mm}}l@{\quad}|@{\quad}c@{\quad}|@{\quad}c@{\quad}|@{\quad}c@{\quad}|@{\quad}c@{\quad}|}
        \hline
        parameter & best fit & 2$\sigma$ & 3$\sigma$ & 4$\sigma$
        \\
        \hline
        $\Delta m^2_{21}\: [10^{-5}\eVq]$
        & 7.9?? & 7.3--8.5 & 7.1--8.9 & 6.8--9.3\\
        $\Delta m^2_{31}\: [10^{-3}\eVq]$
        & 2.2?? & 1.7--2.9 & 1.4--3.3 & 1.1--3.7\\
        $\sin^2\theta_{12}$
        & 0.30? & 0.25--0.34 & 0.23--0.38 & 0.21--0.41\\
        $\sin^2\theta_{23}$
        & 0.50? & 0.38--0.64 & 0.34--0.68 & 0.30--0.72 \\
        $\sin^2\theta_{13}$
        & 0.000 &  $\leq$ 0.031 & $\leq$ 0.051  & $\leq$ 0.073 \\
        \hline
    \end{tabular}
    \caption{ \label{tab:summary-new} 
      Updated version of Tab.~\ref{tab:summary} taking
      into account the background from $^{13}$C in KamLAND: Best-fit values,
      2$\sigma$, 3$\sigma$, and 4$\sigma$ intervals (1 \dof) for the
      three--flavour neutrino oscillation parameters from global data
      including solar, atmospheric, reactor (KamLAND and CHOOZ) and
      accelerator (K2K) experiments.}
\end{table}

The impact of the new background on the bound on $\theta_{13}$ is illustrated
in Fig.~\ref{fig:C-BG-th13}. The left panel shows the $\chi^2$ of KamLAND data
without imposing the constraint $\sin^2\theta_{13} \ge 0$. We observe that the
best fit occurs now closer to the physical region, and the $\Delta\chi^2$
between $\sin^2\theta_{13}=0$ and the best fit point decreases from 3.5 to
1.75. This again shows that the quality of the fit improves. As a consequence
the bound on $\sin^2\theta_{13}$ from KamLAND (constraining the fit to the
physical region) becomes slightly weaker. From the right panel one can see
that the impact on the bound from global data is rather small. Finally, in
Tab.~\ref{tab:summary-new} we update the results of the global three--neutrino
analysis given in Tab.~\ref{tab:summary}, taking into account the new
background in KamLAND data. Only the numbers for $\Dms$ and
$\sin^2\theta_{13}$ change slightly.

\section{June--2006 update}

We update our three--neutrino oscillation parameter analysis using the
most recent data: the first MINOS results using the NuMI Beam which
have been presented at the Neutrino2006 conference~\cite{MINOS-nu06},
the last K2K results~\cite{Ahn:2006zz}, the new update of the Standard
Solar Model (SSM)~\cite{Bahcall:2004pz,Bahcall:2005va} and also latest
measurements from the SNO collaboration~\cite{Aharmim:2005gt}.

\bigskip
\underline{\bf{First MINOS data}}
\smallskip

MINOS is a long--baseline experiment that searches for $\nu_\mu$
disappearance in a neutrino beam with a mean energy of 3~GeV produced
at Fermilab. It consists of a near detector, located at 1 km from the
neutrino source and a far detector located at the Soudan Mine, at
735~km from Fermilab. Recently data corresponding to $1.27\times
10^{20}$~p.o.t.\ have been released~\cite{MINOS-nu06}, slightly more
than the final K2K data sample. In the absence of oscillations
$239\pm17$ $\nu_\mu$ events with $E<10$~GeV are expected, whereas 122
have been observed, which provides a $5.9\sigma$ evidence for
disappearance. In our re-analysis we use spectral data divided into 15
bins in reconstructed neutrino energy, and our allowed region from
MINOS-only is in very good agreement with the official
result~\cite{MINOS-nu06}.  The values of the oscillation parameters
from MINOS are consistent with the ones from K2K, as well as from SK
atmospheric data.

\bigskip
\underline{\bf{Last K2K data}}
\smallskip

Recently the K2K collaboration has published details of the analysis
of their full data sample (K2K-I and K2K-II)~\cite{Ahn:2006zz}. The
data have been taken in the period from June 1999 to November 2004 and
correspond to $0.922\times 10^{20}$~p.o.t. Without oscillations
$158^{+9.2}_{-8.6}$ events are expected whereas only 112 events have
been observed. Out of these, 58 events are single--ring events where
the reconstruction of the neutrino energy is possible, we use these
events to perform a spectral analysis of the K2K data, as described in
Sec.~\ref{sec:k2k-accel-exper}.

\bigskip
\underline{\bf{New SNO-salt data and SSM update}}
\smallskip

We have also updated our solar data analysis in view of the recently
released update of the SSM and the recent SNO data in its salt phase.
Taking into account new radiative opacities, Bahcall et
al.~\cite{Bahcall:2004pz,Bahcall:2005va} obtain new solar neutrino
fluxes, neutrino production distributions and solar density profile.
Among the different solar models they present, we have adopted the one
denoted as BS05(OP), since it agrees better with the helioseismic
measurements. For completeness we have checked also the alternative
model choice BS05(AGS,OP) obtained from recently determined heavy
element abundances. 
The change to this model, however, has a negligible effect on the
result for the oscillation parameters found for the preferred BS05(OP)
option. (For our analysis the main difference between these two models
is the predicted boron flux, which is now more precisely determined by
SNO data).

In addition to the new SSM we include in the present analysis the
latest SNO neutrino flux (CC, NC, ES) measurements given in
Ref.~\cite{Aharmim:2005gt} based on 391 days of data for the SNO
phase~II. With respect to the previous result (254 days) the value of
the CC to NC flux ratio changed from $\phi_{CC}/\phi_{NC} =
0.306\pm0.026\pm0.024$ to the slightly higher value
$\phi_{CC}/\phi_{NC} = 0.340\pm0.023\pm0.030$.

\bigskip
\underline{\bf{Updated three--neutrino analysis}}
\smallskip

\begin{figure}[t] \centering
    \includegraphics[width=.95\linewidth]{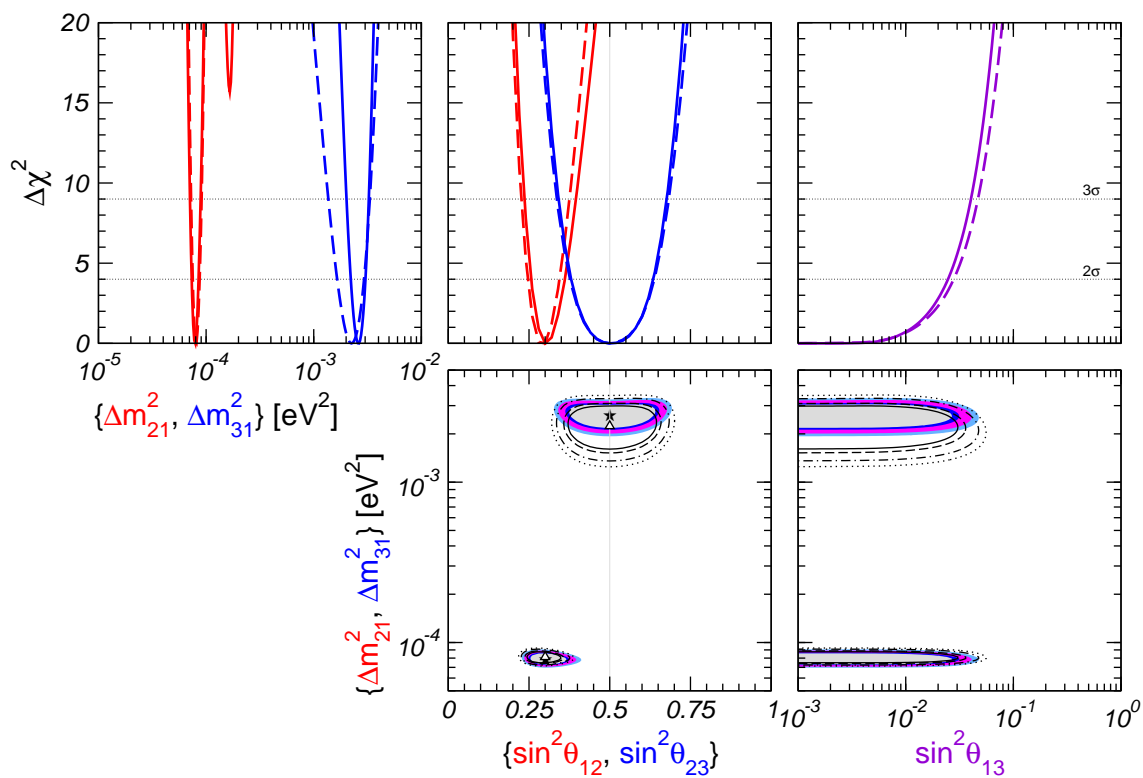}
    \caption{\label{fig:global2} %
      Projections of the allowed regions from the global oscillation
      data at 90\%, 95\%, 99\%, and 3$\sigma$ \CL\ for 2 \dof\ for
      various parameter combinations. Also shown is $\Delta \chi^2$ as
      a function of the oscillation parameters $\sin^2\theta_{12},
      \sin^2\theta_{23}, \sin^2\theta_{13}, \Delta m^2_{21}, \Delta
      m^2_{31}$, minimized with respect to all undisplayed
      parameters. 
      Dashed lines and empty regions correspond to the global analysys
      before this update, while solid lines and coloured regions show
      our most recent results. }
\end{figure}

The results of our updated global three--neutrino analysis are
summarized in Fig.~\ref{fig:global2} and Tab.~\ref{tab:summary2}.
Here we simply highlight the main differences with respect to our
previous results:
\begin{itemize}
\item
An increase in the best fit value for the ``atmospheric'' mass
splitting $\Delta m_{31}^2$ and a narrower allowed range for this
parameter due to the inclusion of the first MINOS data which mainly
lead to a tighter lower bound (see Fig.~\ref{fig:global2}). The
allowed range of $\theta_{23}$ is dominated by atmospheric data as
before, since MINOS provides a very weak constraint on the mixing
angle, similar to K2K.
\item
A small shift in the allowed interval for the ``solar mixing angle''
$\sin^2\theta_{12}$, due to the larger value of last SNO-salt fluxes
compared to its previous measurements and the change in the SSM.  The
so-called high-LMA solution around $\Delta m^2_{21} \approx 2\times
10^{-4}$~eV$^2$ is now slightly less disfavored, with $\Delta\chi^2
\approx 15$.  This follows from the somewhat higher value of
$\phi_{CC}/\phi_{NC}$, as becomes clear from the contours of
$\phi_{CC}/\phi_{NC}$ shown in Fig.~\ref{fig:sol-region}.
\item
A slight improvement in the $\theta_{13}$ bound, mainly because of the
larger value of $\Delta m_{31}^2$ preferred by MINOS data. In the
global analysis also the updates in the solar fit contribute to the
improved bound.
\end{itemize}

\begin{table}[t] \centering
    \catcode`?=\active \def?{\hphantom{0}}
    
\begin{tabular}{|@{\quad}>{\rule[-2mm]{0pt}{6mm}}l@{\quad}|@{\quad}c@{\quad}|@{\quad}c@{\quad}|@{\quad}c@{\quad}|@{\quad}c@{\quad}|}
        \hline
        parameter & best fit & 2$\sigma$ & 3$\sigma$ & 4$\sigma$
        \\
        \hline
        $\Delta m^2_{21}\: [10^{-5}\eVq]$
        & 7.9?? & 7.3--8.5 & 7.1--8.9 & 6.8--9.3\\
        $\Delta m^2_{31}\: [10^{-3}\eVq]$
        & 2.6?? & 2.2--3.0 & 2.0--3.2 & 1.8--3.5\\
        $\sin^2\theta_{12}$
        & 0.30? & 0.26--0.36 & 0.24--0.40 & 0.22--0.44\\
        $\sin^2\theta_{23}$
        & 0.50? & 0.38--0.63 & 0.34--0.68 & 0.31--0.71 \\
        $\sin^2\theta_{13}$
        & 0.000 &  $\leq$ 0.025 & $\leq$ 0.040  & $\leq$ 0.058 \\
        \hline
\end{tabular}
\caption{ \label{tab:summary2} 
  \texttt{2006 updated version of Table 1}.
  Best-fit values, 2$\sigma$,
  3$\sigma$, and 4$\sigma$ intervals (1 \dof) for the
  three--flavour neutrino oscillation parameters from global data
  including solar, atmospheric, reactor (KamLAND and CHOOZ) and
  accelerator (K2K and MINOS) experiments.} 
\end{table}

\bigskip
\underline{\bf{The small parameters: $\alpha$ and $\theta_{13}$}}
\smallskip

\begin{figure}[t]
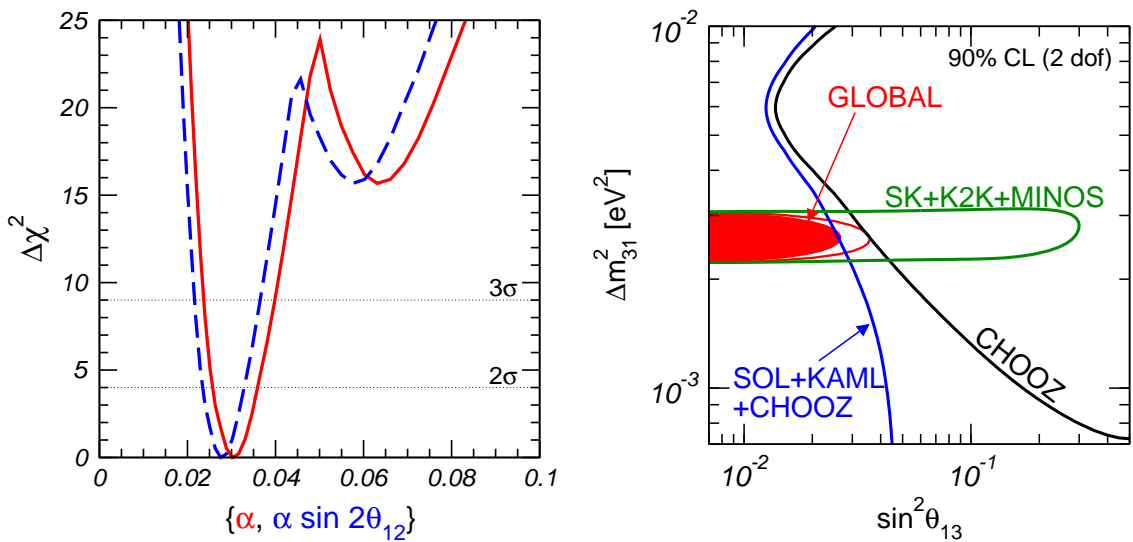
 \centering
    \includegraphics[height=0.3\textheight]{F-fcn.alpha06.eps}\quad
    \includegraphics[height=0.3\textheight]{th13-06.eps}
    \caption{\label{fig:alpha-th13} Left: $\Delta \chi^2$ from global
       oscillation data as a function of $\alpha \equiv \Dms / \Dma$
       and $\alpha\sin 2\theta_{12}$ after the inclusion of MINOS
       data.  Right: 90\% \CL\ upper bound on $\sin^2\theta_{13}$ (2
       \dof) from the combination of all neutrino oscillation data as
       a function of $\Dma$.}
\end{figure}

In Fig.~\ref{fig:alpha-th13} we show the $\chi^2$ as a function of the
$\alpha$ and $\alpha \sin 2\theta_{12}$ parameters relevant for three
flavour effects in future long--baseline experiments, where $\alpha$
denotes the ratio of ``solar'' to ``atmospheric'' mass-squared
differences.  With the new data we obtain the following best fit
values and 3$\sigma$ allowed ranges:
\begin{eqnarray}
    \alpha = 0.030\,, & \quad & 0.024 \le \alpha \le 0.040\,,
    \\
    \alpha \sin 2\theta_{12} = 0.028\,, & \quad &
    0.022 \le \alpha \sin 2\theta_{12} \le 0.037\,.
    \nonumber
\end{eqnarray}
The local minimum at $\Delta\chi^2 \approx 15$ visible in
Fig.~\ref{fig:alpha-th13} appears because of the somewhat weaker
rejection of the high-LMA solution due to the new SNO CC/NC value.

The right panel of Fig.~\ref{fig:alpha-th13} illustrates how the bound
on $\theta_{13}$ emerges from an interplay of the global data. In
particular, the role of solar and KamLAND data is clearly visible,
even after the improvement of the lower bound on $\Delta m^2_{31}$
from MINOS. We find the following bounds at 90\% \CL\ (3$\sigma$) for
1 \dof:
\begin{equation}\label{eq:th13-06}
    \sin^2\theta_{13} \le \left\lbrace \begin{array}{l@{\qquad}l}
    0.033~(0.071) & \text{(solar+KamLAND)} \\
    0.026~(0.054) & \text{(CHOOZ+atmospheric+K2K+MINOS)} \\
    0.020~(0.040) & \text{(global data)}
\end{array} \right.
\end{equation}

\section{September--2007 update}

We have updated our analysis including the new data released by the
MINOS~\cite{Collaboration:2007zz} and KamLAND~\cite{KamLAND:2007}
collaborations. New MINOS data have been collected from June 2006 to
July 2007 (Run-IIa), and they have been analyzed together with the
first data sample (Run-I), with a total exposure of
2.5$\times$10$^{20}$ p.o.t.  In total, 563 $\nu_\mu$ events have been
observed at the far detector, while 738$\pm$30 events were expected
for no oscillation. The most recent data from the KamLAND
experiment~\cite{KamLAND:2007} correspond to a total exposure of 2881
ton-year, almost 4 times larger than 2004 data.  They provide a
very precise measurement of the solar neutrino oscillation parameters,
mainly the mass squared splitting. Apart from the increased statistics
this is also due to the reduction of systematic uncertainties.  Thanks
to the full volume calibration the error on the fiducial mass has been
reduced from 4.7\% to 1.8\%. The main limitation for the $\Delta
m^2_{21}$ measurement comes now from the uncertainty on the energy
scale of 1.5\%. The analysis of the new data is performed in a similar
way as described in appendix A. We use the data binned in equal bins
in $1/E$ to make optimal use of spectral information. As previously we
restrict the analysis to the prompt energy range above 2.6~MeV to
avoid large contributions from geo-neutrinos and backgrounds. In that
energy range 1549 reactor neutrino events and a background of 63
events are expected without oscillations, whereas the observed number
of events is 985.

The Borexino collaboration has also presented their first
data~\cite{Borexino:2007}.  While they provide the first real time
detection of Berilium-7 solar neutrinos, and an important confirmation
of the standard solar model and the large mixing oscillations, they
currently do not affect the determination of neutrino oscillation
parameters.

\begin{figure}[t]
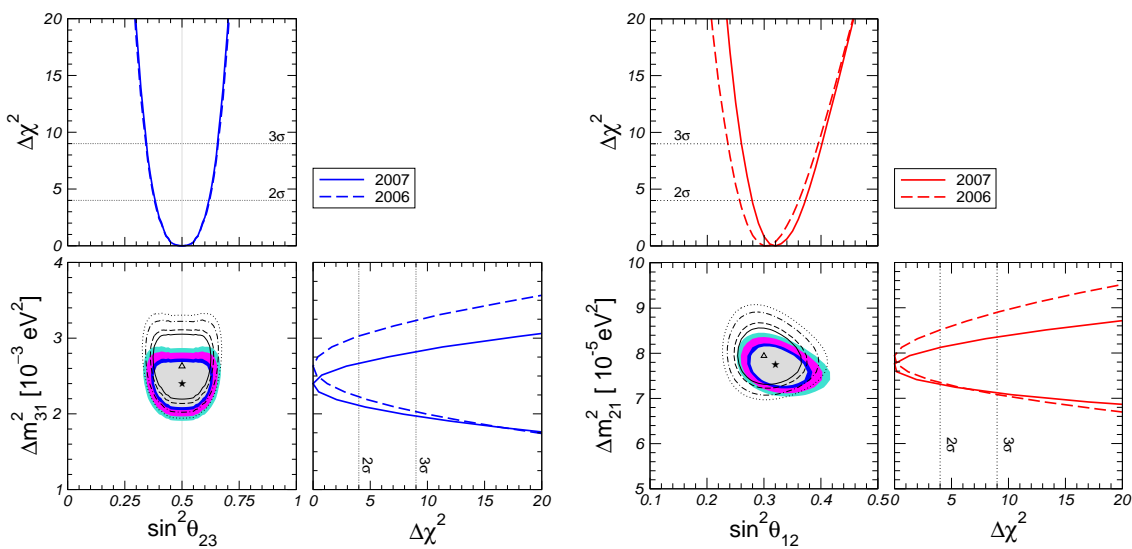
 \centering
  \includegraphics[height=0.3\textheight]{F-atm+LBL-06vs07.eps}\quad
  \includegraphics[height=0.3\textheight]{F-sol+kl-06vs07.eps}
  \caption{\label{fig:last-results} Left: Allowed region in the
    ($\sin^2\theta_{23}$, $\Delta m_{31}^2$) plane before (lines) and
    after (coloured regions) the inclusion of the new MINOS data.
    Right: Allowed region in the ($\sin^2\theta_{12}$, $\Delta
    m_{21}^2$) plane before (lines) and after (coloured regions) the
    inclusion of the new KamLAND data.}
\end{figure}

Our results for the atmospheric and solar neutrino oscillation
parameters are summarized in Fig.\ref{fig:last-results}. In the left
panel we show the allowed region and $\Delta\chi^2$ profiles for the
atmospheric parameters ($\sin^2\theta_{23}$, $\Delta m_{31}^2$) before
and after the inclusion of the MINOS-2007 data. One can appreciate a
decrease in the best fit value and a narrower allowed range for the
``atmospheric'' mass square splitting $\Delta m_{31}^2$. This
illustrates the increasing role of long baseline accelerator data in
the determination of the ``atmospheric'' splitting, a trend that will
get more pronounced in the future. 
The right panel gives the analogous plot for the solar neutrino
oscillation parameters ($\sin^2\theta_{12}$, $\Delta m_{21}^2$), where
one can see how the 2007 KamLAND data provides an improvement in the
determination of the ``solar'' mass square splitting as well as the
lower bound on $\sin^2\theta_{12}$, while the upper bound is still
dominated by SNO solar neutrino data.

\begin{figure}[t]
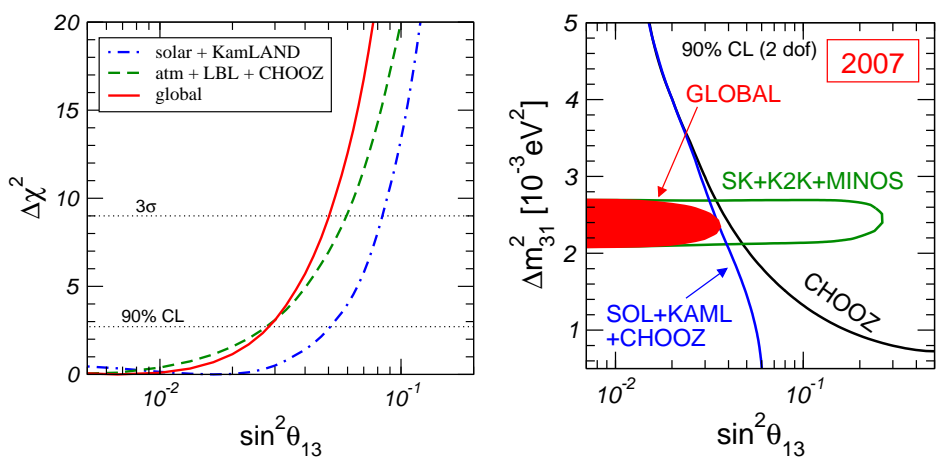
 \centering
  \includegraphics[height=0.25\textheight]{F-th13-2007-tot.eps}\quad
 \includegraphics[height=0.25\textheight]{th13-2007-lin.eps}
  \caption{\label{fig:th13-2007} Left: $\Delta\chi^2$ profiles as a
    function of $\theta_{13}$ from the analysis of different data
    samples.  Right: 90\% \CL\ upper bound on $\sin^2\theta_{13}$ (2
    \dof) from the combination of all neutrino oscillation data as a
    function of $\Dma$.}
\end{figure}

The information concerning the mixing angle $\theta_{13}$ is shown in
Fig.\ref{fig:th13-2007}.  In the left panel we have plotted the
$\Delta\chi^2$ profile as a function of $\sin^2\theta_{13}$ from the
analysis of solar + KamLAND, atmospheric + K2K + MINOS + CHOOZ, and
also from the global analysis of all the data samples.  In the global
analysis we find a slight weakening of the upper bound on
$\sin^2\theta_{13}$ from 0.04 to 0.05 at $3\sigma$. The reason for
this is two-fold. First, the shift of the allowed range for $\Delta
m^2_{31}$ to lower values implies a slightly weaker constraint on
$\sin^2\theta_{13}$, and second, the combination of solar and KamLAND
data prefers a slighlty non-zero value of $\sin^2\theta_{13}$ which,
though not statistically significant, also results in a weaker
constraint in the global fit.
In the right panel we show the 90\% C.L. upper bound on
$\sin^2\theta_{13}$ from the combination of all data samples. No
significant improvement has been otained here, besides the narrower
range for $\Delta m_{31}^2$.

\begin{table}[t] \centering
    \catcode`?=\active \def?{\hphantom{0}}
    
\begin{tabular}{|@{\quad}>{\rule[-2mm]{0pt}{6mm}}l@{\quad}|@{\quad}c@{\quad}|@{\quad}c@{\quad}|@{\quad}c@{\quad}|}
        \hline
        parameter & best fit & 2$\sigma$ & 3$\sigma$ 
        \\
        \hline
        $\Delta m^2_{21}\: [10^{-5}\eVq]$
        & 7.6?? & 7.3--8.1 & 7.1--8.3 \\
        $\Delta m^2_{31}\: [10^{-3}\eVq]$
        & 2.4?? & 2.1--2.7 & 2.0--2.8 \\
        $\sin^2\theta_{12}$
        & 0.32? & 0.28--0.37 & 0.26--0.40\\
        $\sin^2\theta_{23}$
        & 0.50? & 0.38--0.63 & 0.34--0.67\\
        $\sin^2\theta_{13}$
        & 0.007 &  $\leq$ 0.033 & $\leq$ 0.050 \\
        \hline
\end{tabular}
\caption{ \label{tab:summary3} 
  \texttt{2007 updated version of Table 1}.
  Best-fit values, 2$\sigma$ and
  3$\sigma$ intervals (1 \dof) for the
  three--flavour neutrino oscillation parameters from global data
  including solar, atmospheric, reactor (KamLAND and CHOOZ) and
  accelerator (K2K and MINOS) experiments.} 
\end{table}

A summary of the updated neutrino oscillation parameters is given in
Tab.~\ref{tab:summary3}.  Including the 2007 data we find the
following best fit values and 3$\sigma$ allowed ranges for the
parameters characterizing three flavour effects in future
long--baseline experiments, namely $\theta_{13}$, $\alpha$ and $\alpha
\sin 2\theta_{12}$:
\begin{equation}\label{eq:th13-07}
  \sin^2\theta_{13} \le \left\lbrace \begin{array}{l@{\qquad}l}
      0.051~(0.084) & \text{(solar+KamLAND)} \\
      0.028~(0.059) & \text{(CHOOZ+atmospheric+K2K+MINOS)} \\
      0.028~(0.050) & \text{(global data)}
    \end{array} \right.
\end{equation}
\begin{eqnarray}
  \alpha = 0.032\,, & \quad & 0.027 \le \alpha \le 0.040\,,
  \\
  \alpha \sin 2\theta_{12} = 0.029\,, & \quad &
  0.024 \le \alpha \sin 2\theta_{12} \le 0.037\,.
  \nonumber
\end{eqnarray}

\end{appendix}

 

\end{document}